\documentstyle[prl,aps,multicol,psfig,epsf]{revtex}
\renewcommand{\narrowtext}{\begin{multicols}{2} \global\columnwidth20.5pc}
\renewcommand{\widetext}{\end{multicols} \global\columnwidth42.5pc}

\newcommand{\REFrm}{ \def\BIB##1##2##3##4{%
        {{\rm ##1}}\ {\bf ##2}\ (##3),\ ##4}}%
\REFrm
\newcommand{\bold}[1]{\mbox{\boldmath $#1$}}    
\newcommand{\mold}[1]{\mbox{\small\boldmath $#1$}}
\newcommand{\ul}[1]{\underline{#1}}		
\newcommand{\r}{{\bold{r}}}                     
\newcommand{\p}{{\bold{p}}}                     
\newcommand{\q}{{\bold{q}}}                     
\newcommand{\AMD}{{\sl AMD}}			
\newcommand{\FMD}{{\sl FMD}}			
\newcommand{\beq}{\begin{equation}}
\newcommand{\eeq}{\end{equation}}
\newcommand{\beqar}{\begin{eqnarray}}
\newcommand{\eeqar}{\end{eqnarray}}
\newcommand{\Exp}[1]{{\rm e}^{#1}}              

\newcommand{\del}{\partial}
\newcommand{\nuc}[2]{{$^{#2}\hbox{#1}$\ }}
\def\langle{<}
\def\rangle{>}
\newcommand{\VEV}[1]{\langle{#1}\rangle} 
\newcommand{\bra}[1]{\langle{#1}|}
\newcommand{\ket}[1]{|{#1}\rangle}
\newcommand{\SEV}[1]{\prec{#1}\succ}	
\newcommand{\Hml}{{\cal H}}     
\newcommand{\Zcan}{{\cal Z}}                            
\newcommand{\Wei}{{\cal W}}                             
\newcommand{\Z}{\bold{Z}}
\newcommand{\ZC}{\bar{\bold{Z}}}
\newcommand{\W}{\bold{W}}
\newcommand{\WC}{\bar{\bold{W}}}
\newcommand{\rhoE}{\rho_{{\scriptscriptstyle{E}}}}
\newcommand{\C}{\bold{C}}
\newcommand{\F}{\bold{F}}

\newcommand{\z}{\bold{z}}
\newcommand{\zc}{\bar{\bold{z}}}
\newcommand{\E}{{\scriptscriptstyle{E}}}
\newcommand{\FIGCAP}[1]{%
\begin{center}\begin{minipage}{18pc}\begin{small}{#1}\end{small}\end{minipage}\end{center}}
\newcommand{\FIGCAPW}[1]{%
\begin{center}\begin{minipage}{38pc}\begin{small}{#1}\end{small}\end{minipage}\end{center}}
\def\FIGA{%
\begin{figure}	
\begin{center}
\FIGAps
\end{center}
\caption{Occupation probability in a harmonic oscillator.}
\label{fig:Pn}
\FIGCAPW{
The occupation probability for four identical fermions in a harmonic potential
calculated with either distorted states (D: left)
or undistorted states (U: right)
for the observation of the occupation number
and employing the options [Q-$T$], [Cl-$T$], and [Cl-$T'$] in each case.
About $10^4$ states were sampled for each temperature
by means of the Metropolis method.
The exact quantal results are shown by the 
solid dots.
}\end{figure}}

\def\FIGB{%
\begin{figure}	
\FIGBps
\caption{Partition function for a harmonic oscillator.}
\label{fig:Zcan}
\FIGCAP{
The partition function for $A$ distinguishable particles
in a harmonic potential,
calculated either quantally (Q: solid curve) or classically (Cl: dashed curve).
The temperature is expressed in units of the oscillator spacing $D=\hbar\omega$.
}
\end{figure}}

\def\FIGC{%
\begin{figure}	
\FIGCps
\caption{Excitation energy and specific heat of $^{12}$C.}
\label{fig:C12-A}
\FIGCAP{
The mean excitation energy per nucleon $E^*/A$
divided by the temperature $T$ (top)
and the specific heat per particle $C_V/A$ (bottom)
for a canonical ensemble of \nuc{C}{12} nuclei calculated with the \AMD\ model, 
as a function of the temperature $T$.
Squares, triangles and 
circles 
show the results
obtained with various values of the radius constant, $r_0=1.2, 1.5, 2.0$ fm, 
respectively.
Also shown are the results for a free classical gas (dashed lines)
and a nuclear liquid drop model (solid curves)
in which the excitation spectrum contains the known low-energy levels
plus a Fermi-Dirac gas of quasi-particles
with a suitably modified level-density parameter,
$a = A/(8\ {\rm MeV}) ( 1-0.8/A^{1/3} )$ \cite{Fai}.
The marks represent results with three different freeze-out volumes, 
$V=4\pi R^3/3, R=r_0 A^{1/3}$, $r_0$=1.2, 1.5, 2.0 fm.
The sample size for each scenario is $5 \times 10^4$.
In order to give an impression of the sampling error,
the results of two different ensembles are shown for $r_0=2.0$ fm.
}
\end{figure}
}

\def\FIGD{%
\begin{figure}	
\FIGDps
\caption{Specific heat and fragmentation of $^{12}$C.}
\label{fig:C12-B}
\FIGCAP{
The specific heat per nucleon calculated with \AMD\ 
(solid dots)
and with the canonical multifragmentation model
described in the text admitting a maximum of $N$ fragments,
with $N$=2--9 (short dashes).
The radius constant was taken as $r_0=2.0$ fm,
corresponding to a freeze-out density equal to about 20\% of normal.
Also shown are the results for the liquid drop model (solid curve)
and the free nucleon gas (long dashes).}
\end{figure}}

\def\FIGE{%
\begin{figure}	
\FIGEps
\caption{Excitation energy and specific heat of $^{40}$Ca.}
\label{fig:Ca40}
\FIGCAP{
The mean excitation energy per nucleon $E^*/A$
divided by the temperature $T$ (top)
and the specific heat per particle $C_V/A$ (bottom)
for a canonical ensemble of \nuc{Ca}{40} nuclei
calculated with the \AMD\ model,
in a display similar to fig.\ \ref{fig:C12-A}.
The level-density parameter is $A/(8\ {\rm MeV})$.}
\end{figure}}

\def\FIGF{%
\begin{figure}	
\FIGFps
\caption{Ensemble sampling of 
microcanonical temperature.}
\label{fig:MCmetro}
\FIGCAP{
The inverse temperature of distinguishable particles in a harmonic oscillator
as a function of the energy per particle,
as obtained by performing a Metropolis sampling
of the state dependent temperature (\ref{MCbetaE})
using the probability (\ref{MCprobPoi}) for $N=$4 (squares) and 10 (diamonds)
($10^4$ states were sampled for each given energy).
Also shown are the exact analytical results
for the corresponding microcanonical (solid),
canonical (dashed), and classical (dotted) ensembles.}
\end{figure}}

\def\FIGG{%
\begin{figure}	
\FIGGps
\caption{Temporal sampling of 
microcanonical temperature.}
\label{fig:MCsmd}
\FIGCAP{
The inverse temperature of distinguishable particles in a harmonic oscillator
as a function of the energy per particle,
as obtained by performing a sampling over the stochastic time evolution
of a single initial state.
The systems considered and the display
are the same as in fig.\ \ref{fig:MCmetro}
and $10^4$ states were sampled for each given energy as well.}
\end{figure}}

\def\FIGH{%
\begin{figure}	
\FIGHps
\caption{Microcanonical distribution of the Hamiltonian.}
\label{fig:DH}
\FIGCAP{
The distribution of the expectation value of Hamiltonian operator, $\Hml$,
for the systems considered in fig.\ \ref{fig:MCsmd},
for the case with $N$=10 particles,
based on samples of either $10^4$ (squares) or $3\times 10^5$ (diamonds)
consequtive states generated by the proposed stochastic dynamics.
Also shown are the result for the corresponding microcanonical ensembles
(dashed curve)
and that obtained without the phase-space factor $\Hml^{N-1}$
(dotted curve).}
\end{figure}}

\newcommand{\geteps}[1]{\epsfxsize=3in\epsfbox{#1}\vspace*{10pt}}
\def\FIGAps{\begin{minipage}{6.1in}\vspace*{-1cm}~\hspace*{-1cm}%
\epsfxsize=6.0in\epsfbox{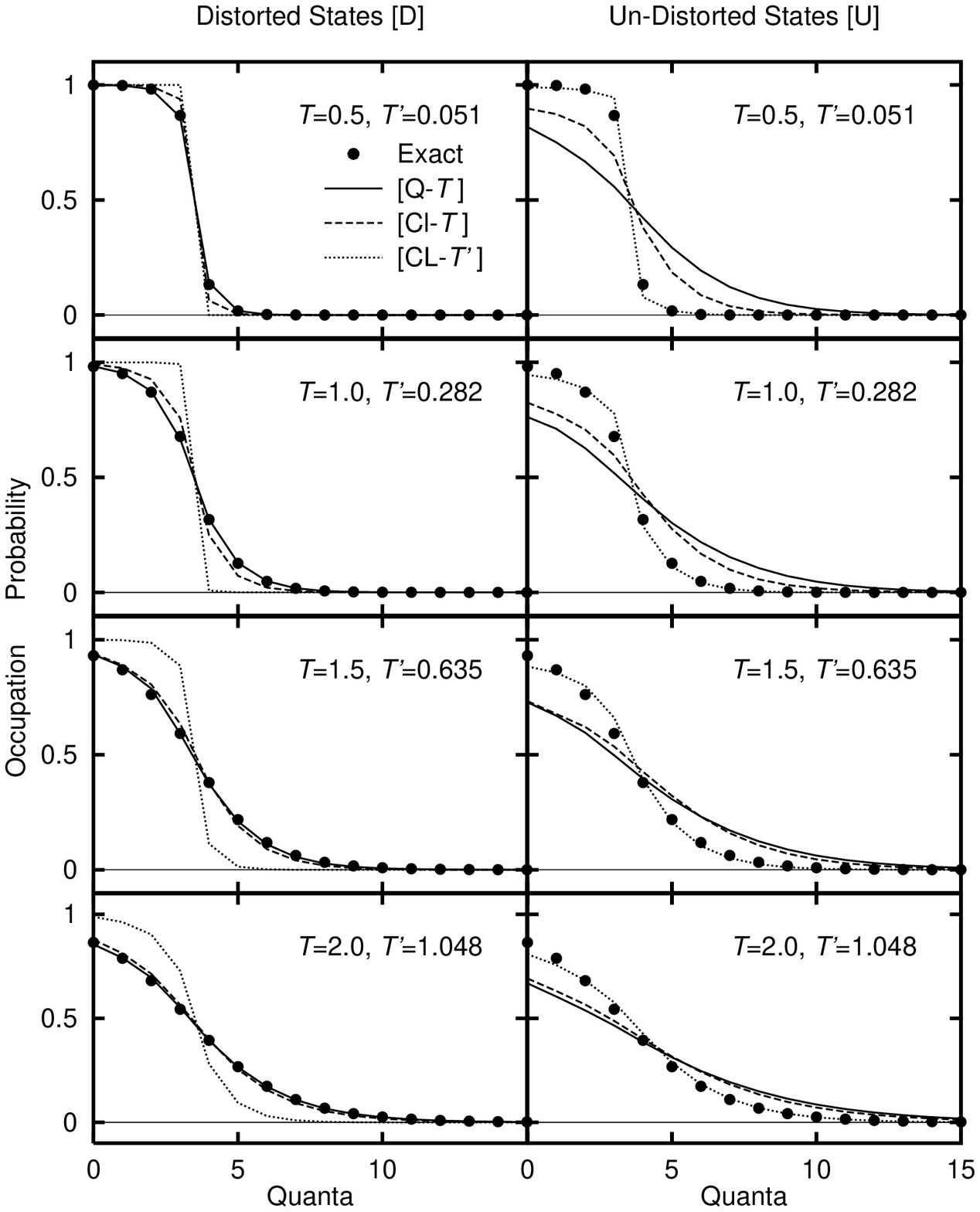}\vspace*{-3.5cm}\end{minipage}}
\def\FIGCps{\begin{minipage}{5.5in}\vspace*{-1.5cm}~\hspace*{-2.2cm}%
\epsfxsize=5.1in\epsfbox{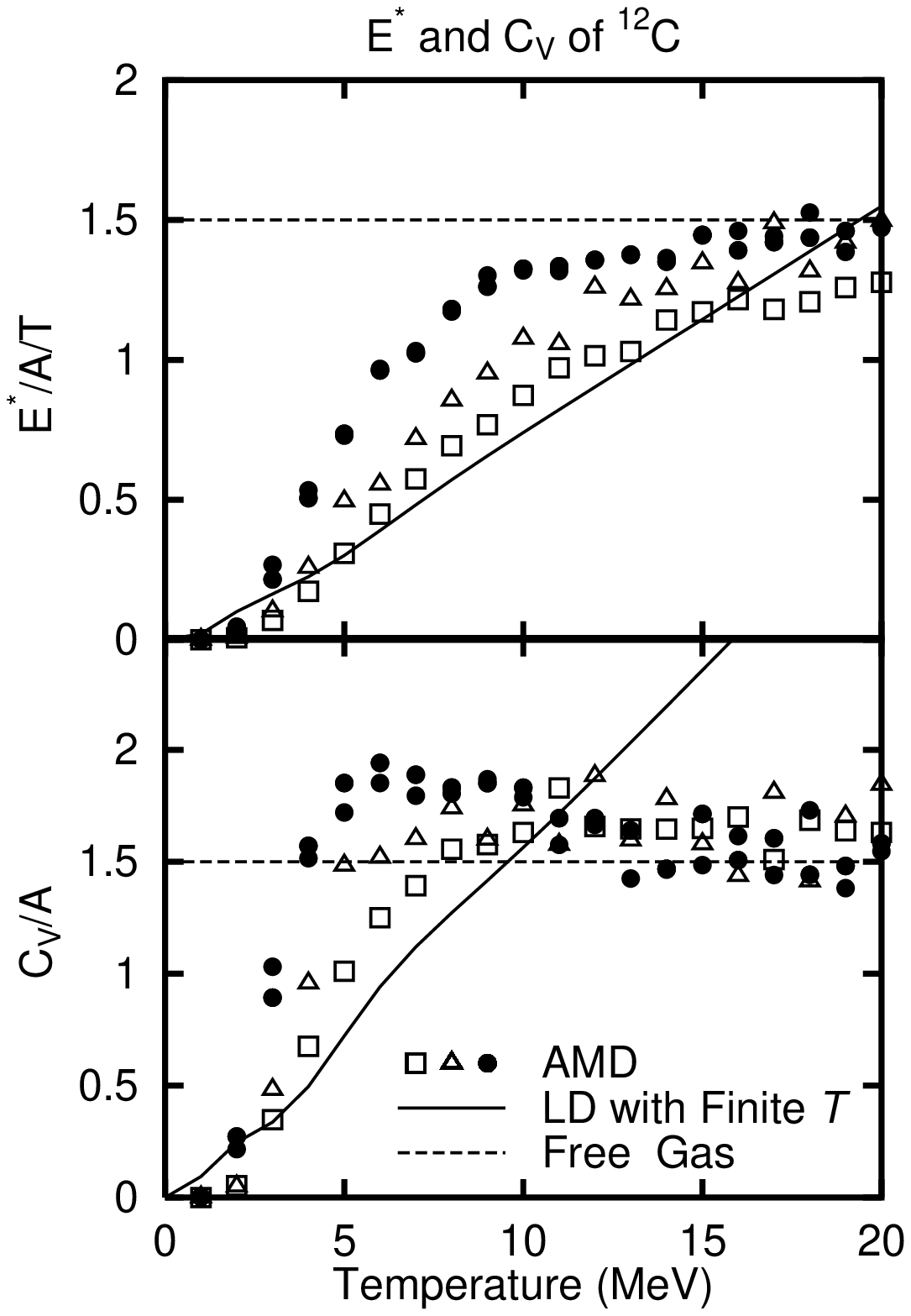}\vspace*{-5.5cm}\end{minipage}}
\def\FIGEps{\begin{minipage}{5.5in}\vspace*{-1.5cm}~\hspace*{-2.2cm}%
\epsfxsize=5.1in\epsfbox{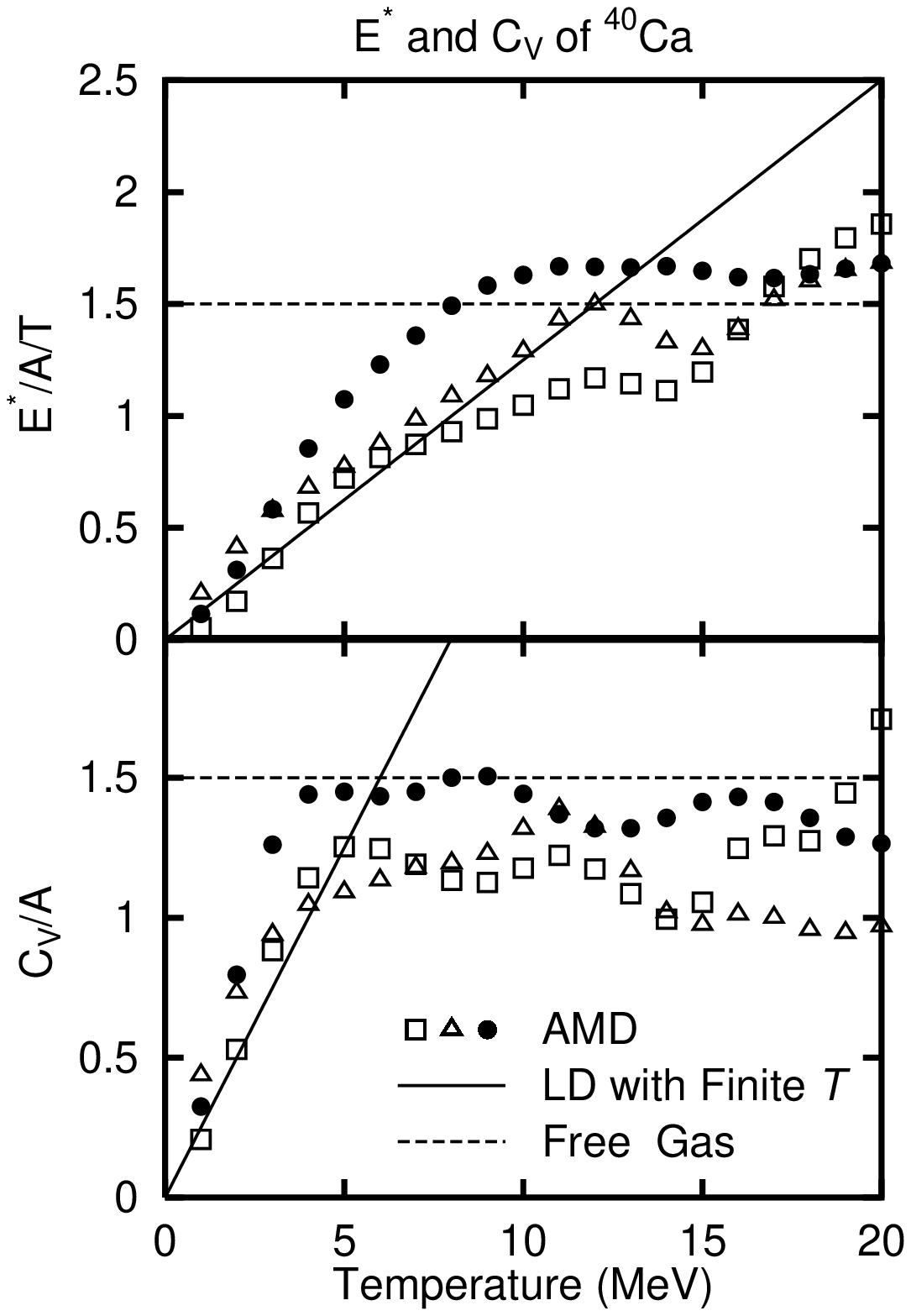}\vspace*{-5.5cm}\end{minipage}}
\def\FIGBps{\geteps{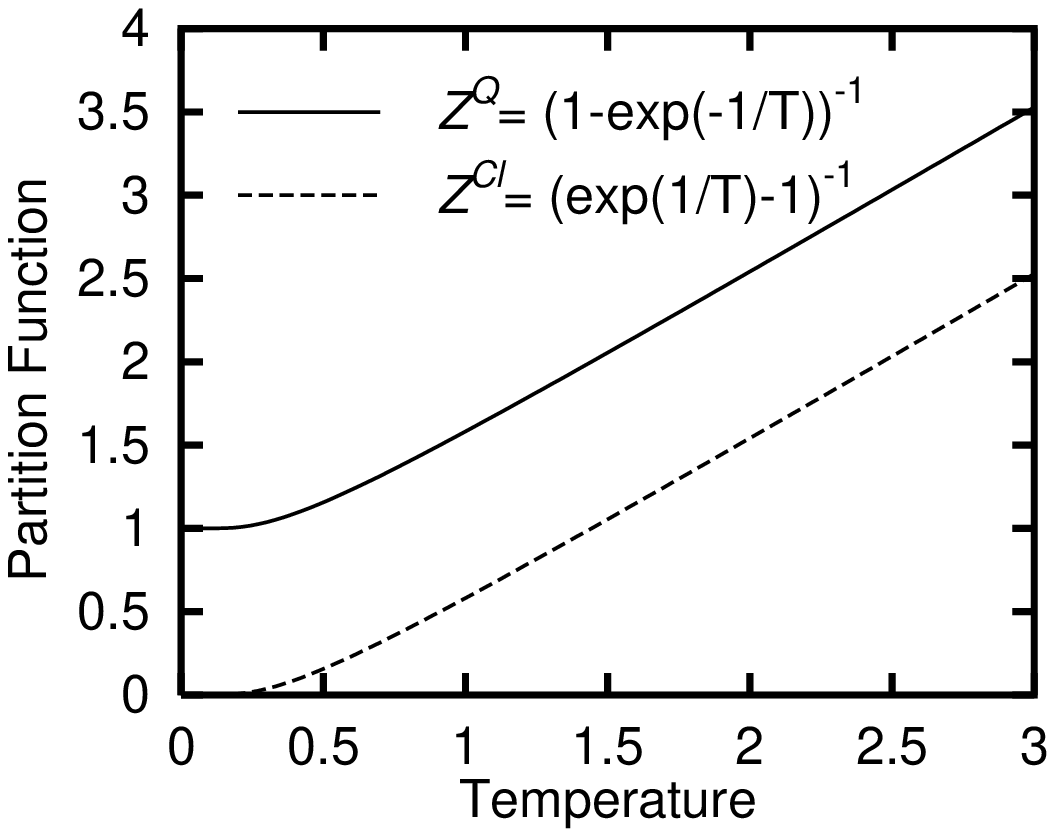}}
\def\FIGDps{\geteps{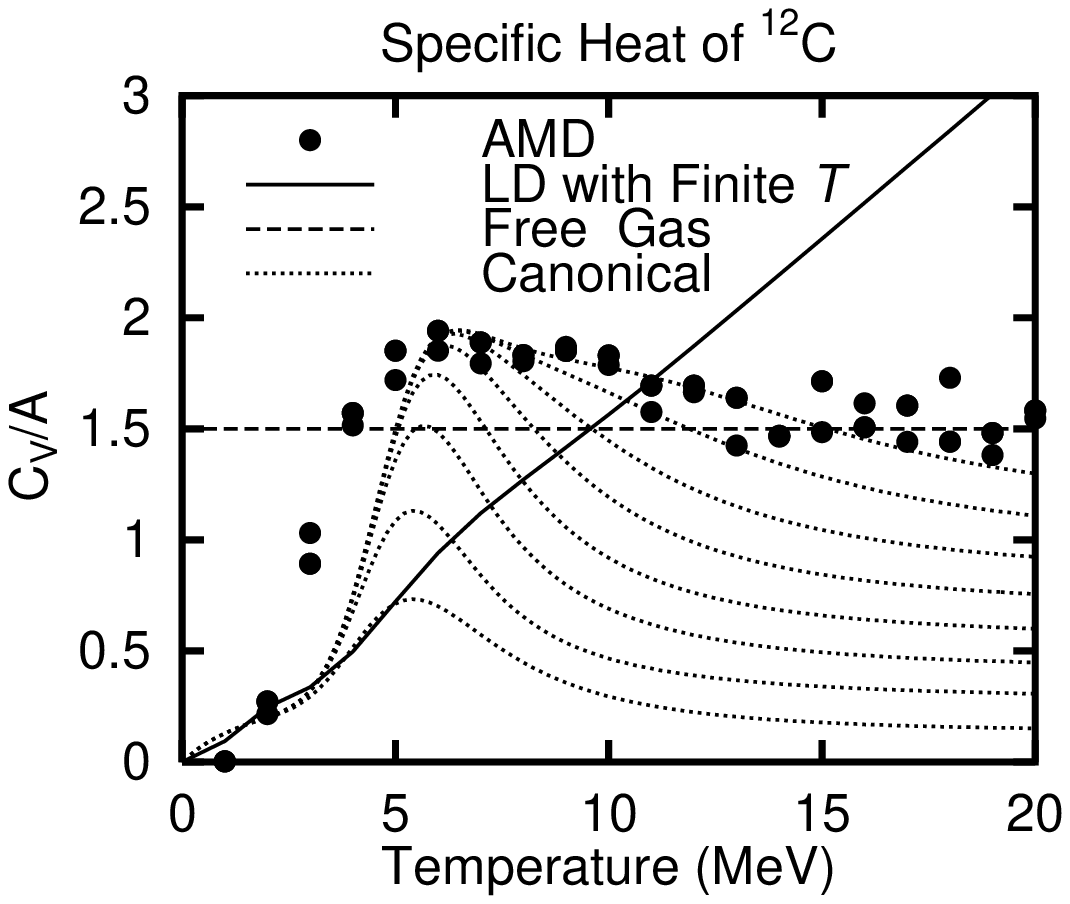}}
\def\FIGFps{\geteps{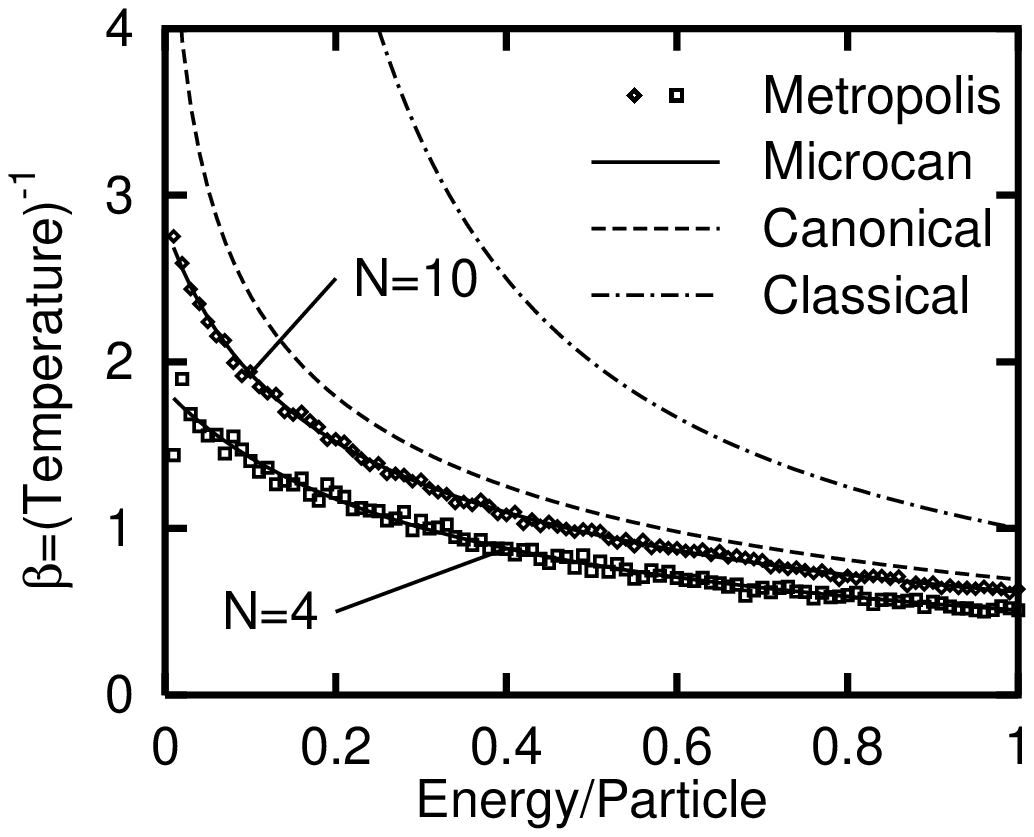}}
\def\FIGGps{\geteps{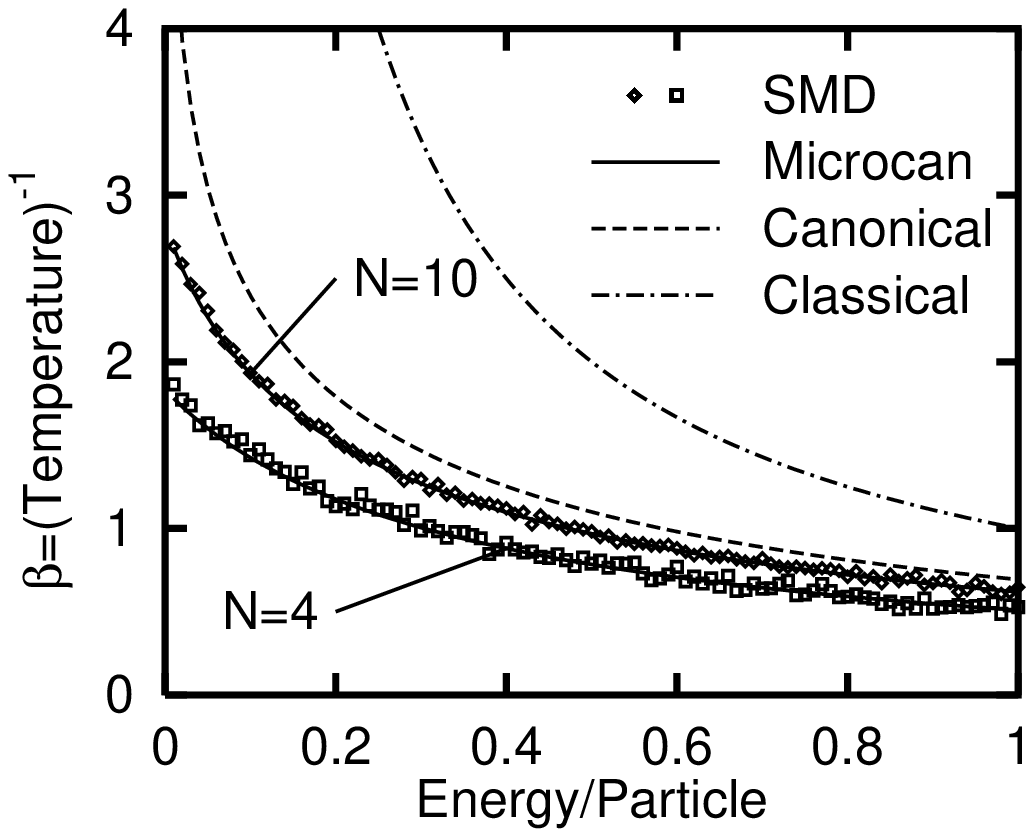}}
\def\FIGHps{\geteps{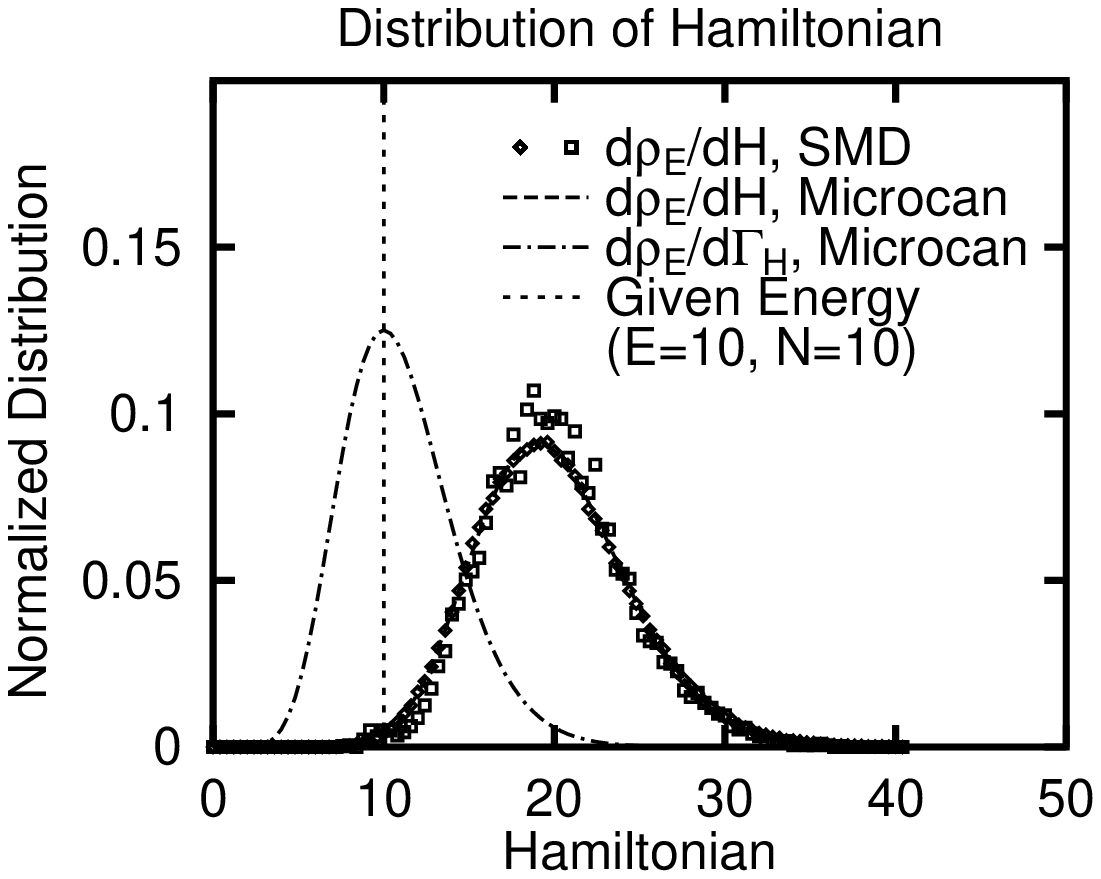}}

\begin{document}


\title{Inclusion of Quantum Fluctuations in Wave Packet Dynamics\footnote{
Submitted to {\sl Annals of Physics}, LBL-38596; HUPS-96-1}}

\author{Akira Ohnishi$^{a,b}$\footnote{
                E-mail: Ohnishi@nucl.phys.hokudai.ac.jp,\,\
                Fax: +81-11-746-5444.}
	\ and \ 
        J\o rgen Randrup$^b$\footnote{
		E-mail: Randrup@LBL.gov,\,\
		Fax: +1-510-486-4794.}
}

\address{$a$.\ 
                Department of Physics, Faculty of Science  \\
                Hokkaido University, Sapporo 060, Japan}
\address{$b$.\ 
                Nuclear Science Division, Lawrence Berkeley Laboratory\\
                University of California, Berkeley, CA 94720, USA}       

\date{14 April, 1996}

\maketitle

\begin{abstract}
We discuss a method by which quantum fluctuations can be included
in microscopic transport models based on wave packets
that are not energy eigenstates.
By including the next-to-leading order term in
the cumulant expansion of the statistical weight,
which corresponds to the wave packets having Poisson energy distributions,
we obtain a much improved global description of the quantum statistical
properties of the many-body system.
In the case of atomic nuclei, exemplified by $^{12}$C and $^{40}$Ca,
the standard liquid-drop results are reproduced at low temperatures and
a phase transformation to a fragment gas occurs as the temperature is raised.
The treatment can be extended to dynamical scenarios
by means of a Langevin force emulating the transitions
between the wave packets.
The general form of the associated transport coefficients is derived
and it is shown that the appropriate microcanonical equilibrium distribution
is achieved in the course of the time evolution.
Finally,
invoking Fermi's golden rule,
we derive specific expressions for the transport coefficients
and verify that they satisfy the fluctuation-dissipation theorem.
\end{abstract}


\narrowtext



\section{Introduction}

Microscopic simulations are employed in various fields of physics
as tools for obtaining both of qualitative and quantitative insight.
In these approaches,
the many-body system is usually represented by a set of classical variables,
even when it is basically quantal in character.
For example,
in molecular dynamics the many-body wave function
is often represented as a (possibly antisymmetrized)
product of parametrized single-particle wave packets,
and the equations of motion for the parameters
are then derived from a suitable variational principle.
As a result,
the quantal features of the system are not fully incorporated
in the description.
In particular,
the system is no longer quantized and the quantum statistical properties
are not properly accounted for.
In effect,
the statistical operator $\exp(-\beta\hat H)$ is being replaced by
the mean-field approximation to the statistical weight,
$\exp(-\beta<\hat H>)$,
although the wave packets are not energy eigenstates.
The wave packet parameters then behave classically rather than quantally.
This shortcoming is especially significant in scenarios
where the statistical properties play a major role,
such as complex processes,
and, consequently,
the quantitative utility of the results needs careful assessment.

This issue is especially relevant in the multifragmentation processes
occurring in energetic collisions between atomic nuclei \cite{GSI,Texas,MSU}.
This phenomenon is of interest since it may provide information on
the liquid-gas phase transition of nuclear matter
that is expected because the nuclear forces
resemble those of a Van der Waals gas.
However, 
interpretation of the data must rely heavily on model simulations, due to the
complexity of the collisions~\cite{BUU,Aichelin,Maruyama,Boal,AMD,AMD2,FMD}.
Since the constituent nucleons are fermions and the associated
quantum statistical features persist to rather high temperatures,
close to the transition temperature from the liquid to the gas phase,
it may then be expected that the fragmentation pattern
is influenced by the basic quantal nature of the nuclear system.


The outcome of complex processes,
such as the fragment multiplicity distribution in nuclear collisions,
is to a large degree governed by the statistical properties of the system,
even if complete equilibrium is not reached.
It is therefore important to understand
to which extent the quantal statistical features
are accounted for in the dynamical model employed.
In our initial study~\cite{Thermal},
we examined the statistical equilibrium properties of 
Antisymmetrized Molecular Dynamics (\AMD)~\cite{AMD}
which describes the time-development of the Slater-determinant of 
single-particle Gaussian wave packets,
and we considered especially non-interacting fermions in one dimension,
either bound in a common harmonic potential 
or moving freely within an interval.
Our principal conclusion was that
the average excitation energy and the specific heat,
considered as functions of the imposed temperature,
generally behave in a classical manner
when the canonical weight employed is consistent
with the dynamics~\cite{Thermal,Ono}.
Furthermore,
we found that the quantal statistical features could be recovered by taking
account of the energy dispersion of each wave packet~\cite{Thermal}.

This result may at first appear somewhat surprising,
since the underlying wave function is antisymmetric
and the quantum statistical properties of the nuclear system
to a large degree arises from the fermionic nature of the nucleons.
Indeed,
that work has generated some debate about the statistical character
of wave packet dynamics,
such as Fermionic Molecular Dynamics (\FMD)~\cite{FMD} and \AMD.
In particular,
Schnack and Feldmeier \cite{SF} (SF)
and Ono and Horiuchi \cite{OH} (OH)
have presented arguments in favor of the statistical properties
of the wave packet systems being quantal rather than classical.
One aim of the present paper is to elucidate this issue.
We shall show that the energy dispersion of each wave packet
plays a decisive role for the appearance of quantal statistical features
and that it can be included in a natural manner
by way of the cumulant expansion of the statistical weight.
Moreover,
the usual molecular dynamics
leads to the classical statistical weight, $\exp(-\beta<\hat H>)$,
while the quantal weight, $<\exp(-\beta\hat H)>$,
can be well approximated by carrying the cumulant expansion
to the next-to-leading order in $\beta$.


The modification of the statistical weight
and the associated distortion of the internal structure of the wave packet
naturally lead to a modification of the dynamics itself.
In our preceding work \cite{PRL},
we have shown that the energy dispersion within each wave packet 
not only leads to the proper quantum statistical properties 
but also gives rise to a Langevin-type random force in the dynamics.
In the field of nuclear physics,
dynamical calculations with fluctuations
have been applied to various dissipative phenomena,
such as nuclear fission \cite{Abe}
and (idealized) nuclear collisions \cite{Randrup,Ayik,Suraud}.
In the first case, the unretained microscopic degrees of freedom
provide a heat bath that induces thermal fluctuations
in the macroscopic degrees of freedom.
In the latter case,
the stochastic nature of the residual two-body collisions
in the Boltzmann-Langevin model produces a statistical ensemble
of possible dynamical histories for the one-body phase-space density.
In contrast to those situations,
the random force considered here has a purely quantal origin,
as it arises from the energy dispersion of the individual wave packets,
and it acts in addition to the statistical fluctuations
induced by the residual scatterings.

The quantal Langevin force arises from the composite nature of each wave packet.
This can be understood from the fact that an energy eigenstate,
which is the observationally relevant entity, 
can be expanded in terms of the wave packets
and therefore instantaneous transitions between those wave packets are possible.
Furthermore, if the system is sufficiently complex to be ergodic,
its time evolution will reflect the associated microcanonical ensemble.
Since the wave packet has a finite energy dispersion,
it may contribute to the phase volume even when its
expectation value differs from the given energy.
In the usual wave packet molecular dynamics,
in which the equations of motion are obtained on the basis of
the time-dependent variational principle,
this kind of relaxation does occur
because the expectation value of the Hamiltonian operator is conserved
and so only the states lying on that hypersurface are dynamically accessible.


The principal aim of the present presentation is two-fold.
The first is to elucidate a microscopic theory
that provides a satisfactory description of
the quantum statistical properties of many-body systems such as nuclei,
including the characteristic quantal behavior at low temperature
and the transition from the quantum liquid to the fragment gas phase.
We will show that this can be achieved 
by employing a simple approximate expression for the exact statistical weight,
$\Wei_\beta = \VEV{\exp(-\beta\hat H)}$,
which is derived by the standard cumulant expansion.
At the same time, we illustrate the importance of the distortion
of the wave packets caused by the canonical operator $\exp(-\beta\hat H)$.
This distortion modifies the expectation value
associated with a given wave packet
and is the key to resolving debate between our work and those of SF and OH.

The second aim is to use the insight gained from static scenarios
to develop a corresponding dynamical theory
in which transitions between states with different energy expectation values
are possible and which relaxes
towards the appropriate quantal microcanonical ensemble.
We will show that this can be achieved by introducing a suitable Langevin force
and employing Fermi's golden rule for specific expressions.
The temperature can be defined without ambiguity and the Einstein relation
between the drift and diffusion coefficients emerges naturally.

This paper is organized as follows. 
In section \ref{canonical} we discuss the statistical properties
of wave packet systems and show the importance of the energy fluctuation.
In section \ref{finite}
we then apply the proposed statistical model to finite nuclei,
specifically \nuc{C}{12} and \nuc{Ca}{40}.
The quantal microcanonical ensemble is then discussed
in section \ref{dynamics}
and we present the dynamical model with quantum fluctuations.
Finally, section \ref{summary} summarizes our studies.

\section{Canonical ensemble of wave packets}
\label{canonical}

In order to set the framework for the discussion,
we start by considering a canonical ensemble
of parametrized many-body wave packets.
The key quantity governing the statistical properties
of a quantal many-body system is the associated partition function,
\beq
\label{Partition1}
\Zcan(\beta) \equiv {\rm Tr}\ {\rm e}^{ -\beta \hat H } \ ,
\eeq
where $\hat H$ is the many-body Hamiltonian operator
and $\beta$ is the inverse
of the temperature of the canonical ensemble considered.

\subsection{Wave packets}

A central issue is how to calculate the statistical properties
on the basis of the parametrized wave packets
commonly employed in microscopic transport simulations.
While our discussion and treatment apply rather generally,
we shall work within the \AMD\ framework,
in order to allow concrete illustrations.
As the basic wave functions
we thus use Slater determinants of Gaussian wave packets,
$(\r|\z)\sim\exp[-\nu(\r-\z/\sqrt{\nu})^2]$ \cite{AMD}.
The $A$-body wave packet $|\Z)$ is then characterized
by its complex centroid parameter vector,
$\Z = (\z_1, \z_2, \ldots , \z_A)$,
and the corresponding normalized wave packet is given by
$\ket{\Z}=|\Z)/\sqrt{(\Z|\Z)}$.

We assume that the wave packets provide a resolution of unity,
as is the case in \AMD.
The anti-symmetrization operator is then represented as
\beq
\label{RoU}
{\cal A}       = \int d\Gamma\ \ket{\Z}\bra{\Z} \ .
\eeq
The operator $\cal A$ projects onto the space of anti-symmetric wave functions,
so it acts as unity within that space. 
In the \AMD\ parametrization, the measure $d\Gamma$ is given by
\beq
d\Gamma(\Z)
        =       \det\C\ d\Z d\ZC
=\det\C\ {d^{3A}{\rm Re}(\Z)\ d^{3A}{\rm Im}(\Z)\over\pi^{3A}}\ .
\eeq
In order to carry out the $6A$-dimensional integrals over $d\Gamma$,
we have adopted a Metropolis sampling method,
which guarantees that the computational effort
is expended in proportion to the relative importance
of the individual wave packets considered.

The action associated with a particular history $\Z(t)$
is given by $S[\Z(t)]=\int dt\ {\cal L}(t)$,
where the Lagrangian is
\beq
{\cal L}(t) = \bra{\Z} i\hbar {\del\over\del t} - \hat H \ket{\Z}\ .
\eeq
The usual demand that the action be stationary
then yields the equation of motion for the parameter vector $\Z$,
\beq
\label{EoM}
{d\Z \over dt} = {i\over\hbar} \F\ ,
\eeq
where the associated generalized force has been employed
\beq
\F=-\C^{-1}\cdot{\del\Hml\over\del\ZC}\ .
\eeq

The elements of the $A\times A$ coefficient matrix $\C$
are spatial tensors with complex elements,
\beq\label{Cij}
C_{ij}  
        = {\del^2 \over \del\zc_i \del\z_j} \log\ (\Z|\Z) \ ,
\eeq
and $\cal H$ denotes the expectation value of the energy,
\beq\label{H}
\Hml
        = \VEV{\Z|\hat H|\Z} = {(\Z|\hat H|\Z) \over (\Z|\Z)}\ .
\eeq

The equation of motion (\ref{EoM}) is entirely classical and,
if it can be brought onto canonical form,
the partition function governing the statistical behavior of the wave packet
parameter $\Z$ is given by the standard classical expression,
\beq\label{Zclass}
\Zcan^{\rm Cl}(\beta)
        =       \int d\Gamma(\Z)\ {\rm e}^{-\beta \Hml(\Z)}
        =       \int d\Gamma(\Z)\ \Wei^{\rm Cl}_\beta(\Z)\ ,
\eeq
where $\Wei^{\rm Cl}_\beta(\Z)=\exp({-\beta \Hml(\Z)})$
is the classical statistical weight of a given wave packet.
Such a transformation has been shown to exist for the
Gaussian wave packets employed in \AMD\ \cite{Thermal},
leading from the wave packet parameter $\Z$
to the phase-space variable $\W=\sqrt{\nu}\q +i\p/2\hbar\sqrt{\nu}$.
The measure is then
\beq
d\Gamma(\Z) = d\W d\WC = \prod_{i=1}^A{d\p d\q \over h^3}\ .
\eeq

The above result (\ref{Zclass}) should be contrasted with
the exact quantal partition function (\ref{Partition1})
which can also be expressed as an integral over the wave packet parameter space,
since the wave packets employed resolve unity (see eq.\ (\ref{RoU})),
\beq
\label{Partition2}
\Zcan(\beta)
	\equiv {\rm Tr}\ {\rm e}^{-\beta \hat H}
     = \int d\Gamma\ \Wei_\beta(\Z)\ .
\eeq
However,
the integrand is more complicated,
\beq
\Wei_\beta(\Z)\
	\equiv\ \VEV{\Z |\ {\rm e}^{-\beta \hat H}| \Z }\ ,
\eeq
depending not merely on the expectation value ${\cal H}\equiv\VEV{\Z|\hat H|\Z}$
but also on the energy fluctuations inherent in the particular wave packet,
as is further discussed below.

\subsection{Statistical weight}

It is a hard task to calculate the exact statistical weight,
$\Wei_\beta(\Z)$, 
since it contains $A$-body operators.
It would in fact be equivalent to solving the dynamics exactly.
It is therefore useful to develop an approximate treatment.

The main problem arises from the energy spread of the wave packets.
Since a given wave packet is generally not an energy eigenstate,
it has a distribution of energy eigenvalues,
as is conveniently described by the associated strength function,
\beq
\rho_\E(\Z)\
	\equiv\ \VEV{\Z|\delta(\hat H - E)|\Z}\
	\not=\ \delta(\Hml - E)\ .
\eeq
Consequently, 
the variance of the energy distribution is generally positive,
\beq
\sigma^2_\E (\Z)\
	\equiv\ \VEV{\Z|\hat H^2|\Z} - \VEV{\Z|\hat H|\Z}^2\ >\ 0\ .
\eeq

When these energy fluctuations are sufficiently small, 
the statistical weight can be evaluated
in the ``mean-field'' or ``static'' approximation,
yielding the classical result,
\beq
\label{WeightST}
\Wei^{\rm Static}_\beta(\Z)
	=	{\rm e}^{-\beta \VEV{\Z|\hat H|\Z}}
	=	{\rm e}^{-\beta \Hml(\Z)}
	=	\Wei^{\rm Cl}_\beta(\Z)\ .
\eeq
The statistical properties are then those of a classical system
with the Hamiltonian function $\Hml(\Z)=\VEV{\Z|\hat H|\Z}$,
{\em i.e.} the same as those exhibited by the wave packet parameter $\Z$,
as discussed above.
This correspondence remains true even for the \AMD\ model,
in which the fermionic nature of nucleons is incorporated
by employing antisymmetrized wave packets \cite{Ono,Thermal}.

We show below how to improve the calculation of the statistical properties
when the energy fluctuations are significant.
For this purpose, 
we first note that the statistical weight can be expressed
as the norm of a wave function that evolves along the imaginary time direction,
\beqar
\label{Cool}
\nonumber
\Wei_\beta(\Z)
        &\equiv& \VEV{\Z|\Exp{-\beta\hat H}|\Z}\
=\       \VEV{\Z|\Exp{-{1\over2}\beta\hat H} \Exp{-{1\over2}\beta\hat H}|\Z}
\\
&=&\	\VEV{\Z({\beta\over2})|\Z({\beta\over2})} \ ,
\eeqar
where the $\beta$ dependent wave packet satisfies
\beq
{\del \over \del \beta} \ket{\Z(\beta)}\ 
        =\     -\hat H \ket{\Z(\beta)}\ ,
\eeq
and, consequently, changes its norm with $\beta$.
The corresponding evolution equation for $\Wei_\beta$ is given by
\beq\label{Wbeta}
{\del \over \del \beta} \Wei_\beta(\Z) 
        = - \Hml_\beta(\Z) \Wei_\beta(\Z)\ ,
\eeq
with the $\beta$ dependent energy expectation given by
\beq
\Hml_\beta(\Z) \
        \equiv \
        { \VEV{\Z(\beta/2)|\hat H|\Z(\beta/2)}
                \over
        \VEV{\Z(\beta/2)|\Z(\beta/2)}
        }\ .
\eeq
The formal solution to eq.\ (\ref{Wbeta}) is
\beq
\label{WeightCool}
\Wei_\beta(\Z)
        = \exp\left[\ - \int_0^\beta\ d\beta'\ \Hml_{\beta'}\ \right] \ .
\eeq
This expression for the statistical weight is exact, 
and when the quantal state $\ket{\Z}$ is an energy eigenstate, 
eq.\ (\ref{WeightCool}) is 
the usual statistical weight $\Wei_\beta=\exp(-\beta E)$, 
where $E$ is the associated eigenvalue.
The difference between (\ref{WeightST}) and (\ref{WeightCool}) 
comes from the fact that the given state $\ket{\Z}$ is not an energy eigenstate
so its energy distribution must be taken into account.

The quantal weight (\ref{WeightCool}) can be calculated
once we know the $\beta$ dependence of the expectation value
of the Hamiltonian operator, $\Hml_\beta(\Z)$.
Generally, $\Hml_\beta(\Z)$ decreases as the temperature is reduced,
since the distortion of the wave packet
caused by the canonical factor $\exp(-\beta E)$ is then more effective.
This feature can also be brought out
by a direct differentiation of $\Hml_\beta$,
\beqar
\nonumber
{\del \over \del\beta}\Hml_\beta(\Z) 
       &=& - \left(
		{
			\VEV{\Z(\beta/2)|\hat H^2|\Z(\beta/2)}
		\over
			\VEV{\Z(\beta/2)|\Z(\beta/2)}
		}
                - \Hml_\beta^2
         \right) \\
        &=& - \sigma^2_\E (\Z(\beta/2))  \leq 0 \ .
\eeqar
Furthermore,
if the given state $\ket{\Z}$ is not orthogonal to the ground state,
as is always true for the generalized coherent states usually employed,
then $\Hml_\beta(\Z)$ converges to the ground state energy $E_{\rm gs}$
as $\beta\to\infty$.

In order to develop a suitable approximation to $\Wei_\beta(\Z)$,
we perform a first-order logarithmic expansion of $\Hml_\beta(\Z)$,
\beq
\log\Hml_\beta(\Z)\ \approx \log\Hml(\Z)\
+\beta{\del\over\del\beta}\log\Hml_\beta(\Z)|_{\beta=0}\ ,
\eeq
and find
\beq
\label{Harmonic}
\Hml_\beta(\Z)\ \approx\ \Hml(\Z)\ {\rm e}^{-\beta D(\Z)}\ .
\eeq
Here the effective level spacing $D$ is equal to the
relative energy variance of the state considered,
\beq\label{D}
D(\Z)   
        \equiv - {\del\over \del\beta} \log \Hml_\beta(\Z) \biggr|_{\beta=0}\,
        = \sigma^2_\E(\Z)/\Hml(\Z)\ .
\eeq
Eq.\ (\ref{WeightCool}) then
yields the corresponding statistical weight,
\beq\label{WeightST1}
\Wei_\beta(\Z)
        \approx \exp\left[ 
                - {\Hml \over D} \left( 1 - \Exp{-\beta D} \right)
                \right] \ , 
\eeq
where the ground-state energy has been subtracted from the Hamiltonian,
${\cal H}(\Z_{\rm gs})=0$.

In order to elucidate this approximation,
we consider a harmonic oscillator with the level spacing $\Delta$
and employ wave packets of coherent form,
$|\Z)=\exp(-\Z a^\dagger)\ket{0}$.
The normalized state is then
\beq
\ket{\Z}=\Exp{-{1\over2}\bar{\Z}\Z}\ \sum_n {\Z^n\over\sqrt{n!}}\ \ket{n}\ ,
\eeq
and its mean energy is $\Hml(\Z)=\bar{\Z}\Z\Delta$.
Moreover, the corresponding spectral distribution is of Poisson form,
\beq\label{rhoa}
\rho_n(\Z)
={1\over n!}\ ({\Hml\over\Delta})^n \ \Exp{-\Hml/\Delta}
={1\over n!}\ (\bar{\Z}\Z)^n \ \Exp{-\bar{\Z}\Z}\ .
\eeq
The $\beta$ evolution of the wave packet $\ket{\Z}$ is easily obtained,
\beq
\ket{\Z(\beta/2)}\ =\
        \Exp{-{1\over2}\bar{\Z}\Z}\ \sum_n\ { \Z^n \over \sqrt{n!}}\ 
                \Exp{-{1\over2}n \beta\Delta}\ \ket{n}\ ,
\eeq 
so, according to eq.\ (\ref{Cool}), the statistical weight then becomes
\beqar
\label{WeightSTa}
\nonumber
\Wei_\beta(\Z)
        &=& \Exp{-\bar{\Z}\Z}\
	\sum_n\ {1\over n!}\ (\bar{\Z}\Z)^{n}\ \Exp{-n\beta\Delta} \\
        &=& \exp\left[ -\bar{\Z}\Z(1-\Exp{-\beta\Delta}) \right]\ .
\eeqar
The $\beta$ dependence of $\Hml$ can then be obtained
by differentiation,
\beqar
\nonumber
\Hml_\beta(\Z) &=& -{\del \over \del\beta} \log\Wei_\beta(\Z)
        = \bar{\Z}\Z\Delta\ \Exp{-\beta\Delta}\ 
\\
	&=& \Hml(\Z)\ \Exp{-\beta\Delta}\ .
\eeqar
Thus the Hamiltonian $\Hml_\beta(\Z)$ exhibits a pure exponential damping.
It then follows that the associated effective level spacing,
$D\equiv-\del\log\Hml/\del\beta$,
is simply equal to the level spacing of the oscillator, $\Delta$.
Furthermore,
we see that the expression (\ref{WeightST1}) for the statistical weight
in fact agrees with the exact result (\ref{WeightSTa}),
with $\Hml/D$ representing the mean number of quanta in the state,
$<n>=\Hml/\Delta=\bar{\Z}\Z$.

The expressions ((\ref{Harmonic}) and (\ref{D}))
are those already used in our previous work \cite{Thermal}.
It was shown there that they yield the proper statistical properties,
in certain exactly soluble cases.
We also note that the statistical weight (\ref{WeightST1})
can be obtained from the general cumulant expansion,
\beqar
\nonumber
\log\Wei_\beta(\Z)
        &=&   -\beta \Hml(\Z) + {1\over 2}\beta^2 \Hml(\Z) D(\Z) + \cdots \\
        &=&   -\beta \Hml(\Z) + {1\over 2}\beta^2 \sigma^2_\E(\Z) + \cdots \ .
\eeqar
The lowest-order term is recognized as the static result,
eq.\ (\ref{WeightST}),
while the inclusion of the next term yields the present
refined treatment, eq.\ (\ref{WeightST1}).
By carrying the cumulant expansion to higher order,
increasingly refined approximations can thus be developed,
if called for.
The approximation adopted here is exact in the case of
distinguishable particles in a harmonic oscillator potential.
Therefore we refer to this treatment as the {\em Harmonic Approximation}.

\subsection{Expectation values}

The effect of the canonical operator $\exp(-\beta \hat H)$ 
on a wave packet is two-fold.
The most obvious effect is the assignment of a statistical weight
for the wave packet as a whole, $\Wei_\beta(\Z)$,
as described above.
However, since each energy component is weighted differently
(namely according to its eigenenergy),
the statistical operator also introduces a distortion
of the internal structure of the wave packet.
This latter effect generally modifies
the expectation value of quantal operators and,
as a consequence,
the evaluation of thermal expectation values is less straightforward
than one might at first have thought.

In general the canonical expectation value
of a quantal operator $\hat O$ is given by 
\beq
\SEV{{\cal O}}_\beta\
	\equiv\	{1\over \Zcan(\beta)}
		{\rm Tr} \left[ \hat O\ \Exp { -\beta \hat H } \right] \ .
\eeq
Using the resolution in terms of the wave packets, eq.\ (\ref{RoU}),
the above expectation value can be written on standard form
as a weighted average,
\beq
\SEV{{\cal O}}_\beta\
	=\ {1 \over \Zcan(\beta)}
		\int d\Gamma(\Z)\ {\cal O}_\beta(\Z)\ \Wei_\beta(\Z)\ ,
\eeq
where the value of the observable
associated with a given wave packet $\ket{\Z}$
is the expectation value of ${\cal O}$
in the {\em distorted} state $\ket{\Z(\beta/2)}$,
\beqar
\label{ObsQ}
\nonumber
{\cal O}_\beta(\Z)\
	&\equiv&\
		\VEV{\hat O}_\beta\
	\equiv\
		{
		\VEV{\Z|\Exp{-\beta\hat H/2} \hat O \Exp{-\beta\hat H/2}|\Z}
		\over
		\VEV{\Z|\ \Exp{-\beta\hat H}|\Z}
		} \\
	&=&	{
		\VEV{\Z(\beta/2)|\hat O|\Z(\beta/2)}
		\over
		\VEV{\Z(\beta/2)|\Z(\beta/2)}
		}
			 \ .
\eeqar

When the temperature is sufficiently high
in comparison with the effective level spacing,
$T\gg D=\sigma_\E^2/\Hml$,
the distortion is very small and we have, 
\beq\label{Oclass}
\SEV{{\cal O}}_\beta^{\rm Cl}\
	=\ {1 \over \Zcan(\beta)}\ 
	\int d\Gamma(\Z)\ \VEV{\Z|\hat O|\Z} \Wei_\beta(\Z) \ .
\eeq
It is relatively easy to exhibit the effect of
the thermal distortion of the wave packets
when the temperature is high.
An expansion of eq.\ (\ref{ObsQ}) shows that the leading correction
to the classical result (\ref{Oclass})
is proportional to the quantal correlation between the observable $\hat O$
and the Hamiltonian $\hat H$,
\beqar
\nonumber
{\cal O}_\beta(\Z)
	&=& {\cal O}(\Z) - \beta 
		\left[
			\VEV{\Z|{1\over2}\{{\hat O},{\hat H}\}|\Z}
			- {\cal O}(\Z)\ {\cal H}(\Z)
		\right]\\
&&		+ \cdots\ ,
\eeqar
where $\{\cdot,\cdot\}$ denotes the anti-commutator.
Thus,
the expectation value of $\hat O$
in the thermally distorted state decreases
if the correlation is positive and conversely.

However,
in cases of practical interest the thermal distortion is not negligible
and must be taken into account in order to obtain quantitatively
useful results \cite{Thermal}.
The effect is well illustrated by the behavior of
the mean energy of the thermal ensemble of wave packets.
Assuming that $\Hml_\beta(\Z)$ exhibits a simple exponential evolution,
as in eq.\ (\ref{Harmonic}), 
the mean energy can be expressed as
\beq
\SEV{\Hml}_\beta\
	=\ {1 \over \Zcan(\beta)}\ 
	\int d\Gamma(\Z)\ \Hml(\Z)\ \Exp{-\beta D} \Wei_\beta(\Z)\ .
\eeq
The distortion factor $\Exp{-\beta D}$ suppresses
the contribution from a given wave packet
(as it should, since it is positively correlated with itself),
and it ensures that the characteristic quantal behavior,
$E^* \propto \Exp{-\beta D}$,
emerges at low temperature, $T\ll D$.

\subsection{Discussion and illustrations}

In our initial work,
we pointed out that the standard \AMD\ model
exhibits classical statistical properties
and suggested that the situation be remedied by
including of the energy fluctuation of each wave packet \cite{Thermal}.
That work has generated some debate about the nature
of the statistical properties of wave packet molecular dynamics,
such as FMD and AMD.
In particular,
Schnack and Feldmeier \cite{SF} (SF)
and Ono and Horiuchi \cite{OH} (OH)
have made the counterclaim that the statistical properties
of the wave packet systems are quantal rather than classical.
Hoping to elucidate the issue,
we now discuss and illustrate the situation in some detail.

In their analyses,
both SF and OH focus on the occupation probability
for the single-particle levels in a harmonic-oscillator potential.
Specifically,
they consider the quantity
\beq\label{Occupation}
P_n(\beta)\ \equiv\  {1 \over \Zcan(\beta)}
		\int d\Gamma(\Z)\ \VEV{\Z|\hat O_n|\Z} \Wei_\beta(\Z) \ ,
\eeq
where ${\hat O}_n=a_n^\dagger a_n$ is the one-body operator
counting the number of particles in the level $n$.
In the work by SF,
the integral over $d\Gamma$ is carried out through
the time-evolution of FMD wave functions in the harmonic oscillator potential
with a perturbative residual interaction;
they have also studied the thermalization between two oscillators
with different frequencies.
The work by OH employs the same wave packets and measure
as our original work \cite{Thermal},
and they have also studied actual nuclear systems
by using the evaporated nucleons as a thermometer.
The results of these studies of the harmonic oscillator
can be summarized as follows.
When the initial energy (SF) or the temperature (OH)
is chosen so as to reproduce the mean energy
of the corresponding quantal canonical ensemble,
then the quantity $P_n$ emerges as being very similar to
the occupation probability of the quantal canonical ensemble.

This result is not unexpected,
since both of \FMD\ and \AMD\
are based on totally anti-symmetrized wave functions
and therefore take account of the Fermi statistics
governing the individual particles.
Moreover,
the relation between temperature and energy in the ensembles
employed by SF and OH is by construction the quantal one.
However, this type of analysis does not address the key issue,
namely the many-body properties of the system.
In order to clarify the situation,
we discuss the following three separate issues:
\begin{enumerate}
\item{\em Statistical weight.}
	The statistical weight $\Wei_\beta$ employed in the integration
	over parameter space can be calculated either quantally (Q) or
	classically (Cl),
	\begin{itemize}
	\item \ul{Quantal} statistical weight:
	\beq
	\nonumber
		\Wei_\beta^{\rm Q}
			= \VEV{\Z | \exp(-\beta \hat H)| \Z }\ .
	\eeq
	\item \ul{Classical} statistical weight:
	\beq
	\nonumber
		\Wei_\beta^{\rm Cl}
			= \exp\left( -\beta\Hml \right) \ . 
	\eeq
	\end{itemize}
\item{\em Expectation value.}
	The value of the observable associated with a given wave packet
	can be calculated with either the thermally distorted states (D)
	or the undistorted states (U),
	\begin{itemize}
	\item Observation with \ul{distorted} states:
	\beq
		{\cal O}^{\rm D}_\beta(\Z)
			=
		{
		\VEV{\Z|\Exp{-\beta\hat H/2} \hat O \Exp{-\beta\hat H/2}|\Z}
		\over
		\VEV{\Z|\Exp{-\beta\hat H}|\Z}
		}
			\equiv {\cal O}_\beta(\Z)
	\eeq
	\item Observation with \ul{undistorted} states:
	\beq
		{\cal O}^{\rm U}_\beta
			= \VEV{\Z|\hat O|\Z}\ 
			\equiv {\cal O}(\Z)
	\eeq
	\end{itemize}
\item{\em Temperature.}
	The temperature can be either taken as the specified one ($T$)
	or readjusted so the expectation value of the energy
	is matched exactly ($T'$),
	\begin{itemize}
	\item $T$: The given temperature $T=1/\beta$
	is the one used for the statistical weight and the observation.
	\item $T'$: The given temperature $T$ is regarded as spurious
		and is replaced by $T'$ which is obtained by demanding	
		that the mean energy of the system match the exact value.
	\end{itemize}
\end{enumerate}

The exact calculation discussed  in the previous section
then corresponds to the option [Q-D-$T$],
whereas the calculation by OH corresponds to [Cl-U-$T'$].
Although SF perform the ensemble sampling by means of the time evolution,
their residual interaction ensures ergodicity
and the treatment by SF is then be essentially the same as that by OH.

To illustrate the effect of the different treatments,
we consider four fermions in a harmonic oscillator
and calculate the occupation probability of each single particle level,
for various temperatures.
Figure \ref{fig:Pn} shows the results
based on either distorted (D) or undistorted (U) wave packets
and employing the options [Q-$T$], [Cl-$T$], and [Cl-$T'$] in each case.
The quantal statistical weights were calculated in the harmonic approximation,
and the quantal observation is carried out
by assuming the $\beta$ evolution of ${\z_i}$ to be 
$\z_i(\beta/2) = \z_i(0) \exp(-\beta\hbar\omega/2)$.

\widetext
\FIGA
\narrowtext

Comparison with the exact result (solid dots) shows that
both [Q-D-$T$] and [Cl-U-$T'$] reproduce the exact results very well,
which is in accordance with the reports by SF and OH.
Even though the latter treatment is not formally correct,
this agreement is not unexpected,
as we shall now discuss.
The key to achieving a correspondence between the two treatments
is to regard the distorted state $\ket{\Z(\beta/2)}$
as the state $\ket{\Z'}$ sampled by SF and OH. 
Then, in the harmonic approximation where $\Hml_\beta=\Hml\exp(-\beta D)$, 
the quantal statistical weight $\Wei_\beta^{\rm Q}$ can be rewritten
in terms of a modified inverse temperature $\beta'$
and the expectation value with respect to $\ket{\Z'}$,
\beqar
\nonumber
\log \Wei^{\rm Q}_\beta(\Z)
	&\equiv& - {\Hml(\Z) \over D} \left( 1 - \Exp{-\beta D}\right)\ 
\\
\nonumber
	&=&\ - {\Hml(\Z')\ \Exp{\beta D} \over D}
		\left( 1 - \Exp{-\beta D}\right) \\ 
	&=& - \beta'\ \Hml(\Z')\
	\equiv\	\log \Wei_{\beta'}^{\rm Cl}(\Z')\ ,
\eeqar
where the modified temperature is determined by
\beq
\beta' = {\Exp{\beta D} - 1 \over D}
= \beta(1+{1\over2}\beta D +\dots)\ .
\eeq
Consequently,
the quantal partition function can be expressed
in terms of the classical statistical weight for the distorted state,
\beq\label{Z'}
\Zcan^{\rm Q}(\beta)
	\equiv	\int d\Gamma(\Z)\ \Wei^{\rm Q}_\beta(\Z)
	=	\int J\ d\Gamma(\Z')\ \Wei^{\rm Cl}_{\beta'}(\Z')\ ,
\eeq
where $J$ is the Jacobian associated with the transformation from
$\Z$ to $\Z'$,
\beq
J\ \equiv\
\left| \begin{array}{cc} \W & \WC \\ \W' & \WC' \end{array}\right|\
=\	{\det{\C} \over \det{\C'}}\
\left| \begin{array}{cc} \Z & \ZC \\ \Z' & \ZC' \end{array}\right|\ .
\eeq
Comparing the above reformulation (\ref{Z'})
with the classical partition function,
\beq
\Zcan^{\rm Cl}(\beta') = \int d\Gamma(\Z')\ \Wei^{\rm Cl}_{\beta'}(\Z')\ ,
\eeq
we see that if the Jacobian $J$ is independent of $\Z$
then the two partition functions are proportional,
\beq
\Zcan^{\rm Q}(\beta) = J\ \Zcan^{\rm Cl}(\beta')\ , \\
\eeq
In addition, 
the expectation value ${\cal O}_\beta(\Z)$
is then the same as $\VEV{\Z'|\hat O|\Z'}$ by definition.
Consequently,
the thermal averages of the observable $\hat O$ are also equal,
\beqar
\nonumber
\SEV{{\cal O}}^{\rm Q}_\beta
	&\equiv&	{1 \over \Zcan^{\rm Q}(\beta)}
	\int d\Gamma(\Z)\ {\cal O}^{\rm Q}_\beta(\Z)\ \Wei^{\rm Q}_\beta(\Z)\\ 
\nonumber
	&=&	{1 \over J\ \Zcan^{\rm Cl}(\beta)}\ 
		J \int d\Gamma(\Z')\ {\cal O}^{\rm Cl}_{\beta'}\  
			\Wei^{\rm Cl}_{\beta'}(\Z')\
\\
	&=&	\SEV{{\cal O}}^{\rm Cl}_{\beta'}\ .
\eeqar
This result brings out very clearly how it is possible,
for each given temperature $T$,
to introduce a modified temperature $T'$
so that the quantal result is reproduced.

However,
this kind of treatment leads to inconsistent thermodynamics.
The mean energy $\SEV{\Hml}$ should be given by
$ - \del \log \Zcan / \del \beta$,
and that is indeed the case in the approximate quantal treatment
of ref.\ \cite{Thermal}.
But the modified treatment in [Cl-U-$T'$] yields the mean energy
\beq
\SEV{\Hml}^{\rm Cl}_\beta\
	=\	{1\over \Zcan^{\rm Cl}(\beta')}
		\int d\Gamma(\Z')\ \Hml(\Z')\ \Wei^{\rm Cl}_{\beta'}(\Z')\ ,
\eeq
while a differentiation of the partition function (\ref{Z'}) leads to
\beq
-{\del \over \del \beta} \log\Zcan^{\rm Cl}_{\beta}
	=	{1\over \Zcan^{\rm Cl}(\beta)}
		\int d\Gamma(\Z')\ \Hml(\Z')
		\ \Exp{\beta D}\ \Wei^{\rm Cl}_{\beta'}(\Z') \ ,
\eeq
which contains the additional factor $d\beta'/d\beta = \exp(\beta D)$.
Therefore, the modification of the temperature parameter
renders the thermodynamics inconsistent.

A further difference between the two treatments
is the partition function itself.
The classical partition function $\Zcan^{\rm Cl}$ deviates from
the quantal partition function $\Zcan^{\rm Q}$ by the Jacobian factor $J$.
If the determinant ratio $\det\C/\det\C$ is ignored,
we have $J \simeq \exp(-A\beta D)$.	
This ideal situation is realized in the case of $A$ distinguishable
particles in a harmonic oscillator potential,
where the level spacing $D=\hbar\omega$ is a constant.
Furthermore,
the Jacobian $J$ is independent of $\Z$ but still depends on $\beta$,
of course.
The partition functions $\Zcan^{\rm Q}$ and $\Zcan^{\rm Cl}$
are shown in fig.\ \ref{fig:Zcan}.
While the quantal partition function approaches unity when $T \to 0$,
the classical partition function tends to zero.
This occurs because the phase space for $\Z'$ 
is drastically reduced relative to the original phase space of $\Z$
when the temperature is smaller than the level spacing,
as already discussed.
The fact that the phase space, and hence also the statistical fluctuations,
are too small in the ensembles employed by SF and OH
may present a practically important shortcoming of those treatments,
since the partition function is the key quantity governing
the fragment size distribution in actual nuclear collision processes.

\FIGB

\section{Statistical properties of finite nuclei}
\label{finite}

In the previous section,
we employed a few simple soluble cases to illustrate the general treatment.
We now turn to more realistic scenarios
and consider the statistical properties of finite nuclei.

As we have seen,
we need to determine the $\beta$ evolution of the wave packet.
In the harmonic approximation,
this is equivalent to calculating the energy dispersion $\sigma^2_\E(\Z)$
of the wave packet,
since level spacing $D=\sigma^2_\E/\Hml$
is the key quantity determining the statistical weight.
More generally,
the equations of motion (EOM) for the wave packet parameters $\Z$
can be employed for this task
by replacing the real time $t$ by the imaginary time $-i\hbar\beta$.
%
In the \AMD\ model,
\beq
\label{Cooling}
{\del \Z \over \del t} = {i\over\hbar} \F\phantom{mm}
        \stackrel{t\to-i\hbar\beta}{\rightarrow}\phantom{mm}
{\del \Z \over \del \beta} =\F\ .
\eeq
The first equation governs the usual time development,
while the second one can be used for the estimation of the statistical weight.
It may be noted that the second equation
is usually employed for the construction of the ground-state nuclei,
such as those used as initial conditions in nucleus-nucleus collisions,
since it effectively cools the system down.
It is therefore refered to as the ``cooling'' equation \cite{AMD}.

In the present work,
we have adopted eq.\ (\ref{WeightST1}) as the statistical weight,
\beq
\Wei_\beta(\Z) = \exp\left[ 
                - {\Hml \over D} \left( 1 - \Exp{-\beta D} \right)
                \right] \ ,
\eeq
and the effective level spacing
associated with a given wave packet $\ket{\Z}$
is calculated by the \AMD\ cooling equation,
\beqar
\nonumber
D(\Z)\ 
	&=& \ -{\del\over\del\beta}\log{\cal H}_\beta\biggr|_{\beta=0}\
\\
	&=& \ -{1\over 2 \cal H}
	\left[ 
		{\del{\cal H}\over\del\Z}{\del\Z\over\del\beta}\
	+	{\del{\cal H}\over\del\ZC}{\del\ZC\over\del\beta}\
	\right]
=\bar{\F} \cdot {{\bold{C}} \over \Hml} \cdot \F\ .
\eeqar
Working in the harmonic approximation where $\Hml_\beta = \Hml\exp(-\beta D)$,
we wish to calculate the temperature dependence of the
mean energy $\prec\Hml\succ_\beta$ and the specific heat,
\beqar
\nonumber
C_V(\beta)\
	&\equiv&\
	-\beta^2 {\del^2\over\del\beta^2}\log\Z(\beta)\
\\
	&=&\ \beta^2 \left[
			\SEV{\VEV{\hat H^2}_\beta}_\beta - \SEV{\Hml}^2
		\right]\ ,
\eeqar
where the expectation value of ${\hat H}^2$ is given by
\beq
\VEV{\hat H^2}_\beta\  =\ D\ \Hml(\Z)\ \Exp{-\beta D}\
	=\	\sigma^2_\E(\Z)\ \Exp{-\beta D}\ .
\eeq
The harmonic approximation demands that $D$ is constant
along the $\beta$ evolution described by the cooling equation.
This is not strictly true for real nuclei,
since the energy surface is not exactly quadratic.
However, at low excitations
a quadratic behavior holds approximately
since the first derivative of $\Hml$ vanishes in the ground state. 
Moreover,
at high temperatures where $\beta$ expansion is valid,
the quantum fluctuaton effects are well described by the cumulant expansion
and the harmonic approximation is then reliable in this regime as well.
Therefore we expect the harmonic approximation
to provide a reasonable starting point. 

\FIGC

Using the above framework, we have studied the statistical 
properties of \nuc{C}{12} and \nuc{Ca}{40} with the \AMD\ model. 
The effective nuclear interaction used is Volkov No.\ 1 \cite{Volkov}
and the width parameter $\nu$ and the Majorana mixture are chosen
so as to fit the binding energy and the {\em r.m.s.} radius.
The phase space integral are evaluated by means of a Metropolis sampling method.

Figure \ref{fig:C12-A} shows the excitation energy $E^*$
and the specific heat $C_V$ of \nuc{C}{12} as a function of the temperature.
For convenience,
the excitation energy has been divided by the temperature
(the free value is then a constant ($=3/2$)).
For comparison,
we also show $E^*$ and $C_V$ for free classical particles (dashed line)
and for a finite-temperature liquid drop model
that includes known low-energy nuclear levels
augmented by a Fermi-Dirac gas (solid curve).

The calculated results exhibit the expected behavior:
At low temperature the excitation energy behaves quadratically,
$E^*/A \approx aT^2$, 
and as the temperature increases it approaches
its free value, $E^*_{\rm free}=3T/2$.
This behavior is consistent with the transition
from the liquid phase (or quantal phase) to the gas phase (or classical phase).
The same features appear for the specific heat:
It grows approximately linearly at low temperatures, $C_V/A \approx 2aT$, 
and then approaches its free value,
$C_V^{\rm free}/A=3/2$.
In addition,
the low-temperature results obtained for normal density, $r_0=1.2$ fm,
are very close to those of the standard liquid drop model.
These results are very satisfactory,
since they yield a good description of the statistical properties
of finite nucleus over the entire range of temperatures. 

\FIGD

It is interesting to study the volume dependence of these quantities, 
since the freeze-out volume is the most important parameter
in the statistical models of multifragmentation, 
{\em i.e.}, it determines the ratio between binary
and multifragment decays \cite{Gross}.
The qualitative behavior
(the transition from the liquid-like phase to the gas-like phase)
does not change with volume,
but it becomes easier to excite the system when the volume grows larger.
This is because the excitation of intrinsic modes in a single nucleus
is being overwhelmed by the agitation of multi-fragment configurations.
In addition,
there appears a region in which the specific heat exceeds its free value of 3/2,
reflecting the rapid activation of many degrees of freedom 
that were effectively frozen at low temperatures.
Thus,
such a rapid increase of the specific heat
may signal the onset of multifragmentation.

In order to explore the validity of this picture,
we have performed a simple statistical calculation on the fragment yields.
The system is assumed to appear as a configuration
of distinct exitable nuclear fragments, $\{n_i\}$,
where $n_i$ is the multiplicity of the nuclear species $i$.
The total partition function is then given by
\beq
\Zcan	=	{1\over\Zcan_0} \sum_{\{n_i\}}\ 
		\prod_{i}\ 
		{1\over n_i!}\ \Zcan_i^{n_i}\ 
		\Exp{-\beta V(\{n_i\})}\ ,
\eeq
where $\Zcan_i$ is the contribution to partition function
arising from a fragment of type $i$ within the volume $\Omega$,
\beq
\Zcan_i	= \Omega \left( {M_i T \over 2\pi\hbar^2} \right)^{3/2} \zeta_i \ ,
\eeq
and $\Zcan_0$ refers to the compound nucleus $^AZ$.
The intrinsic degrees of freedom are taken into account
through the effective intrinsic partition function $\zeta_i$
which contains the ground-state degeneracy $g_i=2J_i^{g.s.}+1$
and the excited levels up to $E^*=E_0 + \Delta\ (A_i-4)$
for fragments with $A_i > 4$,
where $E_0$ is the minimum energy for particle evaporation and $\Delta=2$ MeV,
as suggested in ref.\ \cite{Fai}. 
Furthermore,
the potential energy includes the binding energy difference 
and the inter-fragment Coulomb potential, 
\beq
V(\{n_i\})	= B(A,Z)	- \sum_\alpha B(A_\alpha,Z_\alpha)
		+ \sum_{\alpha < \beta} V_{\alpha\beta}   \ ,
\eeq
where $\alpha$ and $\beta$ enumerate the fragments
in the particular configuration $\{n_i\}$.
The fragment binding energies $B$ are taken from experiment
and the Coulomb interaction is fixed by the barrier height.
Figure \ref{fig:C12-B} shows the resulting the specific heat per nucleon 
calculated with this fragment canonical model
with up to nine-fragment configurations included,
as well as the \AMD\ results from fig.\ \ref{fig:C12-A} with $r_0=2.0$ fm.
It is seen that the \AMD\ behavior is well reproduced
when a sufficient fragment multiplicity is admitted
in the statistical calculation.
It is especially noteworthy that the reproduction of
the rapid increase at $T \sim 4$ MeV,
requires the inclusion of channels with four or five fragments.
This illustration shows that multifragmentation appears naturally
as the volume is increased and, moreover,
the model presented in this work can describe this phenomenon.

It is necessary to study statistical properties of heavier nuclei 
in order to fully elucidate the relation between multifragmentation
and the liquid-gas phase transition,
since the latter is closely related to the properties of nuclear matter.
Figure \ref{fig:Ca40} shows the calculated excitation energy
and specific heat for \nuc{Ca}{40}.
Since the sampling number at each temperature and volume is 
only around $2\times 10^3$, the statistical error is not small.
However, the qualitative features are almost the same as for \nuc{C}{12}
and, in particular,
the low-temperature correspondence with the finite-temperature
nuclear liquid-drop model is good.

\FIGE

\section{Dynamics with quantal fluctuations}
\label{dynamics}

Until now we have studied the statistical properties
of an assembly of nucleons.
Below we briefly describe how the insights gained
can be applied in dynamical scenarios.
The most important result of the previous analyzes is
that the quantal nature of the system demands that
the energy dispersion of each wave packet be taken into account.
This follows from the fact that the wave packet is not an eigenstate
of the Hamiltonian operator and therefore the observed energy
exhibits fluctuations around the given initial energy value \cite{PRL}.
This feature then affects the dynamical evolution significantly,
as can be most easily seen from the expression for
the microcanonical phase volume, as shown below.
Since the long-time dynamical evolution of a closed system
should reflect the associated microcanonical ensemble,
we first discuss the statistical properties of a microcanonical ensemble.

\subsection{Quantal microcanonical ensemble}
\label{micro}

For an ergodic system,
in which each state is reachable from any given initial state,
the long-time dynamical evolution should reflect
the associated microcanonical ensemble.
Therefore,
the statistical properties of a dynamical model
are related more closely to the microcanonical ensemble
than to the canonical ensemble.
The microcanonical phase volume is given by
\beqar
\nonumber
\Omega(E) 
        \equiv {\rm Tr}\left[ \delta(\hat H - E) \right]
      &=& \int d\Gamma\ {\VEV{\Z|\delta(\hat H - E)|\Z} \over \VEV{\Z|\Z} }
\\
\label{MCa}
       &=&	\int d\Gamma\ \rhoE(\Z) \ .
\eeqar
Since the wave packets are not energy eigenstates, {\em i.e.} 
$\rhoE(\Z)\not=  \delta(\Hml-E)$,
those states with $\Hml \not= E$ can also contribute 
to the microcanonical phase volume
according to the probability of $\rhoE(\Z)$.
It is interesting to note that resolution of unity (\ref{RoU})
causes the strength function $\rhoE(\Z)$ plays a dual role:
While it is defined as the energy distribution
in a {\em single} wave packet $\Z$,
it also expressed the relative weight of {\em different} wave packets
in the microcanonical ensemble characterized by the energy $E$.

From the microcanonical phase volume (\ref{MCa}),
the ensemble temperature $T$ can be defined unambiguously
by means of the formal thermodynamical relation
as the inverse of the rate at which the phase volume increases with energy,
\beqar
\nonumber
{1\over T}
	&\equiv& {\del \over \del E}\ \log\Omega(E)
	=	{1\over\Omega(E)} \int\ d\Gamma\ {\del \rhoE(\Z) \over \del E}
\\
\label{MCbeta}
	&=&	{1\over\Omega(E)} \int\ d\Gamma\ \beta_\E(\Z)\ \rhoE(\Z)\ ,
\eeqar
where the contribution from a given wave packet is given by
\beq
\label{MCbetaE}
\beta_\E(\Z)
	\equiv {\del \over \del E}\ \log\rhoE(\Z)\ .
\eeq
We note that this latter state-dependent quantity can have either sign.
(Since $\rhoE(\Z)$ is largest when $E\approx\Hml$,
it will generally increase with $E$ when $E<\Hml$
and decrease when $E>\Hml$.)

The microcanonical phase volume (\ref{MCa}) is intimately related to
the partition function of the associated canonical ensemble,
\beqar
\nonumber
\Zcan(\beta)
        &\equiv& \int_0^\infty dE\ \Omega(E)\ \Exp{-\beta E}
        =     \int d\Gamma \int_0^\infty dE\ 
                \rhoE(\Z) \Exp{-\beta E}
\\
        &=&     \int d\Gamma\ \Wei_\beta(\Z) \ .
\eeqar
Thus statistical properties of both
the microcanonical and the canonical ensemble
can be studied on the same footing 
through $\rhoE(\Z)$ and $\Wei_\beta(\Z)$.
We also note that when the energy eigenvalue distribution 
of the given state $\ket{\Z}$ is of Poisson form,
we recover the same statistical weight $\Wei_\beta(\Z)$
as found earlier (eq.\ (\ref{Harmonic})).

\FIGF

In order to illustrate the situation,
we consider a microcanonical ensemble of
$N$ distinguishable particles in a common harmonic oscillator
and calculate the corresponding inverse ensemble temperature
by means of eq.\ (\ref{MCbeta}).
The energy distribution $\rhoE(\Z)$ is assumed to be 
a continuous Poisson distribution,
\beq
\label{MCprobPoi}
\rhoE(\Z)\ \propto\	\Exp{-\Hml}\ {\Hml^E \over E!}\
	=\	\Exp{-\Hml}\ {\Hml^E \over \Gamma(E+1)}\ ,
\eeq
using units in which $\hbar\omega=1$.
(The number of states for a given energy is then
$\Omega(E)=\Gamma(E+N)/[\Gamma(E+1)\Gamma(N)]$.)
Using the above expression as the weight function,
it is then possible to employ the Metropolis sampling technique
to calculate the average in eq.\ (\ref{MCbeta}),
$\prec \beta_E \succ$.
The result is shown in fig.\ \ref{fig:MCmetro}
and leads to a perfect reproduction of the exact inverse temperature
for the microcanonical ensemble,
as one should expect.
The temperature depends on the number of particles $N$,
even though the particles are distinguishable,
and may be compared with the results for
the corresponding canonical and classical ensembles.

\subsection{Quantal Langevin force}

It follows from the previous discussion
that after the system has undergone a sufficiently long time evolution
the probability distribution $\phi(\Z)$ for finding the state $\ket{\Z}$ 
should be proportional to the corresponding level density,
\beq
\label{MCDist}
\phi(\Z)\phantom{m} \stackrel{t\to\infty}{\to}\phantom{m} \rhoE(\Z)
= \Exp{-{\cal F}(\Z)} \ ,
\eeq
where
\beq
{\cal F}  \equiv -\log\rhoE(\Z)\
\eeq
plays the role of an effective potential for the parameter $\Z$.
The evolution of the wave packet parameter vector, $\Z(t)$,
as determined by the equation of motion (\ref{EoM}),
is entirely deterministic,
without any physical effect of the spectral structure of the wave packet.
Consequently, as we have discussed above,
the system does not relax towards
the appropriate quantum statistical equilibrium.
Furthermore,
this malady is inherent in the classical treatment
and it could not be remedied by merely including a collision term
in the standard manner,
since the evolving wave packet would then still remain on the
equi-Hamiltonian surface $\Hml=E$ and, therefore,
the distribution $\phi(\Z)$ would not relax to
the microcanonical  ensemble (\ref{MCDist}). 

In order to provide the system with an opportunity for exploring
and exploiting the various eigencomponents contributing to its wave packet,
we wish to augment the equation of motion (\ref{EoM}) by a stochastic term
that may cause occasional transitions between different wave packets. 
The analogy of the distribution (\ref{MCDist})
with the classical Boltzmann distribution $\Wei_\beta(\Z)=\exp(-\beta\Hml)$
suggests a way to introduce such a stochastic term,
since the latter distribution can be obtained
by introducing a Langevin term in the equation of motion.
Analogously,
the quantal distribution can be obtained by augmenting the
deterministic equation of motion (\ref{EoM}) by a Langevin term \cite{Risken},
\beq
\label{LangevinA}
\dot{\Z}\ =\	\bold{h}\ +\ \bold{g} \cdot \bold{\zeta} \ ,
\eeq
where the dot over $\Z$ indicates the rate of change over and above
that given by eq.\ (\ref{EoM}) and
the random complex vector $\bold{\zeta}$ represents white noise,
\beq
\SEV{\bar{\bold{\zeta}}_n(t)\ \bold{\zeta}_{n'}(t')}\
	=\	2\ \delta_{nn'}\ \delta(t-t')\ .
\eeq

In order to see that the Langevin equation (\ref{LangevinA})
indeed can produce the desired equilibrium distribution (\ref{MCDist}),
it is helpful to consider the corresponding Fokker-Planck equation,
\beqar
\nonumber
& &\dot{\phi}(\z_1,\cdots,\z_A;t)\ 
\\
&=&\ 
\label{FP}
	\left[ 	-\sum_{n=1}^A {\del\over\del\z_n} V_n
	+\ \sum_{nn'}^A {\del^2\over\del\z_n\del\zc_{n'}}M_{nn'}  \,
	+ {\rm c.c.}\, \right] \phi\ ,
\eeqar
The transport coefficients $\bold{V}$ and $\bold{M}$
govern the early growth rate of the average value of the parameter vector
and the associated covariance tensor, respectively,
starting from a given value $\Z$.
Therefore, the diffusion coefficient is given by
$\bold{M}=\bold{g}\cdot\bold{g}$.
Furthermore,
if we demand that the quantal distribution (\ref{MCDist}),
$\phi\sim\exp(-{\cal F})$,
be an equilibrium solution to this equation,
the form of the drift coefficient follows,
\beq
V_n\ =\  -\sum_{n'} M_{nn'} {\del{\cal F}\over\del\bar{\z}_{n'}}\
	+\ \sum_{n'} {\del\over\del\bar{\z}_{n'}}M_{nn'}\ .
\eeq
The second term represents the noise-induced drift
which enters in the general case when the tensor $\bold{M}$
depends on $\Z$ \cite{Risken}.
Since this dependence is often unimportant,
we shall ignore this term in the following, for convenience.

Since $\cal F$ is predominantly determined by $\Hml$,
its derivative is proportional to the driving force $\del\Hml/\del\ZC$
and the drift coefficient can then be rewritten,
\beq\label{Einstein}
\bold{V}\
\approx\ -\bold{M}\cdot{\del {\cal F} \over\del\ZC}\
\approx\ -\beta_\Hml\ \bold{M	}\cdot{\del\Hml \over \del\ZC}\ .
\eeq
We have here introduced the inverse state-dependent canonical temperature,
\beq
\beta_\Hml(\Z)\	\equiv\
		{\del{\cal F} \over \del\Hml}\ ,
\eeq
and the relation (\ref{Einstein}) is recognized
as a manifestation of the Einstein relation.

The two state-dependent temperatures introduced,
$\beta_E\equiv-\del{\cal F}/\del E$
and $\beta_\Hml\equiv\del{\cal F}/\del\Hml$,
are identical if $\rhoE(\Z)$ depends on $E$ and $\Z$
only through either $\Hml-E$ or $\Hml/E$.
This is approximately the case in most cases of interest
and they can therefore be regarded as being very similar.
For example,
for the distinguishable particles considered in sect.\ \ref{micro},
with the strength function given by (\ref{MCprobPoi}),
we have
\beqar
\beta_\Hml
	&=&	{\Hml - E \over \Hml}\ ,\\
\nonumber
\beta_\E
	&=&	\log(\Hml)-{d\over dE}\log\Gamma(E+1)\
	\approx\ \log(\Hml/E)
\\
	&=&	\beta_\Hml + {1\over 2}\beta_\Hml^2 + \cdots\ .
\eeqar

\FIGG

In order to illustrate the utility of the stochastic treatment,
we show in fig.\ \ref{fig:MCsmd} 
the inverse ensemble temperature as obtained
with the proposed Langevin equation (\ref{LangevinA}).
The systems are the same as those considered in fig.\ \ref{fig:MCmetro}
and the integral over states is calculated 
by averaging over the configurations obtained at $10^4$ time steps,
starting from a single random initial wave packet.
The results demonstrate that the quantal microcanonical distribution
can be realized as a result of the stochastic pseudodynamics
produced by eq.\ (\ref{LangevinA}).
This fact can be also seen from the resulting distribution of
energy expectation values $P(\Hml)$,
as is illustrated in fig.\ \ref{fig:DH} for $N$=10.
Since the probability for finding the system with a given value of $\Z$
is proportional to $\rhoE(\Z)$,
the Hamiltonian distribution should be given by
\beq
P(\Hml)\ d\Hml\\ =\
\rhoE(\Z)\ {d\Gamma\over d\Hml}\ d\Hml\  \sim\ \rhoE(\Z)\ \Hml^{N-1}\ d\Hml\ .
\eeq
The results indicate that this is indeed the case.

Although the stochastic term appearing here
is somewhat similar to the usual random force
that arises from the interaction between the particles under consideration
and the heat bath in which they are embedded,
it is important to recognize that the present random force
arises from the energy dispersion of each wave packet
and thus has a purely quantal origin.
Therefore,
we refer to this random force as the {\sl quantal Langevin force}.
We show below how this quantal Langevin force can be implemented
once the interaction between the particles has been specified.

\FIGH

\subsection{Characterization of the quantal Langevin force}
\label{VD}

In the preceding,
we have introduced the Langevin equation
and shown that it leads to the desired quantal equilibrium distribution.
However,
in order to apply the treatment to actual dynamical processes,
it is necessary to determine the specific properties
of the transport coefficients.
For this purpose,
we seek guidance from Fermi's golden rule
and assume that the differential rate of transitions from a given wave packet
$\ket{\Z}$ to others near $\ket{\Z'}$ is of the following form,
\beq\label{Fermi}
\label{rate}
w(\Z\rightarrow\Z')\ 
	=\ {2\pi\over\hbar}\ |<\Z'_\E|\hat{V}|\Z_\E>|^2\ \rhoE(\Z')\ .
\eeq
Here the operator $\hat V$ represents a suitable ``residual" interaction
and $E$ is a specified energy which is usually taken as the expectation 
value of the originally specified initial state.
The utility of the above form (\ref{Fermi}) is underscored by the fact that
it leads to the correct microcanonical equilibrium distribution.
This can be verified by considering the equilibrium condition,
\beq
\phi(\Z) \ w(\Z \to \Z') = \phi(\Z') \ w(\Z' \to \Z)\ ,
\eeq
which is evidently satisfied when $\phi(\Z)\sim\rho_E(\Z)$,
due to the symmetry of the matrix element in (\ref{rate}).

In analogy with the distortion occurring for the wave packets
in a canonical ensemble (see sect.\ \ref{canonical})),
the initial and final states appearing in the above matrix element,
$\ket{\Z_\E}$ and $\ket{\Z'_\E}$,
are related to the actual states, $\ket{\Z}$ and $\ket{\Z'}$,
by an appropriate distortion 
arising from the microcanonical constraint on the dynamical system.
In principle,
they should be the  respective components of the wave packets
associated with the energy eigenvalue $E$.
However, since the projection to energy eigenstates poses a hard task,
we have adopted the canonical distortion as a suitable approximation,
\beq
\ket{\Z_\E}\ \approx\
	{\ket{\Z(\beta_\Hml/2)} \over \sqrt{ \Wei_\beta(\Z)} }\
	=\ {\Exp{-\beta_\Hml\hat H/2} \ket{\Z} \over 
	\sqrt{\VEV{\Z|\Exp{-\beta_\Hml\hat H}|\Z}}} \ .
\eeq
Application of the cooling equation (\ref{Cooling})
then yields the parameter characterizing the distorted wave packet,
\beq
\Z_\E	= \Z + {1\over2} \beta_\Hml\ \F\ .
\eeq
It is easily seen that the energy expectation value
of the distorted wave packet $\ket{\Z_E}$ is equal to $E$,
through first order in $\beta_\Hml$.

Simple approximate expressions for the moments of $w(\Z\to\Z')$
can be derived by expanding the transition rate (\ref{rate}) around $\Z_\E$,
where both the matrix element and the energy spectrum have the largest values
(see Appendix \ref{App}).
Then the transition rate from $\Z$ to a specific final state $\Z'$ is given by
\beqar
\nonumber
w(\Z\to\Z')\ 
	&\approx&\
	{2\pi\over\hbar}
	| {\cal V} |^2\ \rhoE(\Z_\E)\ 
\\
\label{w}
	&\times&
	\Exp{
	-\left(\delta\bar{\mold{Z}}-{1\over2}\beta_\Hml\bar{\mold{F}}\right)
		 \cdot \C  \cdot
		\left(\delta\mold{Z}-{1\over2}\beta_\Hml\mold{F}\right)
	}\ ,
\eeqar
where $\delta\Z=\Z'-\Z$ and
${\cal V}\equiv<\Z_\E|\hat{V}|\Z_\E>$
denotes the expectation value of the interaction in the initial state $\Z$.
%
%
The total rate of transitions
from the given state $\Z$ into any final state $\Z'$
can now be readily calculated,
\beq
\label{w0}
w_0(\Z)\
	\equiv\ \int d\Gamma'\ w(\Z\to\Z')\
	\approx\ {2\pi\over\hbar}\ |{\cal V}|^2
		\rhoE(\Z_\E)\ .
\eeq
%
The expected number of transitions taking place during a small time interval
$\Delta t$ is then $n_0=w_0 \Delta t$,
which may also be interpreted as the probability
that a transition occurs during $\Delta t$.
Therefore this total rate $w_0$ can be regarded as 
the inverse lifetime of the wave packet $\ket{\Z}$.

Since the transport coefficients $\bold{V}$ and $\bold{M}$
govern the early growth rate of the average value
of the parameter vector $\Z$
and the associated covariance tensor, respectively,
they can be obtained from the moments of the microscopic transition rate,
\beqar
\nonumber
V_n	&\equiv&\
	\int d\Gamma'\ \delta\z_n\ w(\Z\to\Z')\
\\
\label{Vn}
	&\approx&\ 
-w_0\beta_\Hml \left(\bold{C}^{-1}\cdot{\del{\cal H}\over\del\ZC}\right)_n
=w_0 \beta_\Hml\F	\ ,		\\
M_{nn'}
	&\equiv&\
	\int d\Gamma'\ \delta\z_n\ \delta\bar{\z}_{n'}\ w(\Z\to\Z')\
\\
\label{Dnn}
	&\approx&\ w_0\ (\bold{C}^{-1})_{nn'}\ ,
\eeqar
to leading order in $\beta_\Hml$.
We note that the Einstein relation (\ref{Einstein})
is indeed satisfied by these expressions and, moreover,
the drift coefficient is proportional to the generalized force $\F$,
as would be expected.
It may also be remarked that both the center-of-mass position
and the total momentum remain unchanged on the average,
since $\sum_n V_n=0$,
while the individual dynamical evolutions
will exhibit diffusive Brownian-type excursions around those averages.
This behavior is to be expected,
since the the energy $\cal H$ is no longer a constant of motion
but will fluctuate around the specified value $E$.

The above simple approximate expressions
(\ref{w0}), (\ref{Vn}), and (\ref{Dnn})
make it possible to simulate the Langevin evolution
by picking the stochastic changes $\delta\z_n$ in accordance with
the transport coefficients $\bold{V}$ and $\bold{M}$. 
It is here important that the overlap matrix $\bold{C}$ is positive definite
so its square root exists.
Thus it is fairly straightforward to implement
the proposed stochastic treatment.

\section{Summary}
\label{summary}

In the present work,
we have addressed the treatment of quantum fluctuations
in microscopic descriptions based on wave packets,
a problem encountered in a broad range of fields involving quantum physics.
Since the wave packets are not energy eigenstates,
the statistical operator $\exp(-\beta\hat H)$ cannot be treated
as a $c$-number.
In order to take account of the associated spectral distribution
and ensure that the statistical properties are quantal,
it is necessary to introduce suitable modifications
relative to the ordinary equations of motion for the wave packet parameters
which are basically classical in character,
having been derived from the time-dependent variational principle.

We have formulated a simple but apparently successful treatment
by including the first correction term in the cumulant expansion of the
statistical weight.
The associated small parameter is $\sigma_E^2/T\Hml$,
where $\Hml\equiv\VEV{\Z|\hat H|\Z}$ is the mean energy of a wave packet
and $\sigma_E^2$ is the corresponding energy variance.
This approach is exact when the spectral distribution is of Poisson form,
as is often the case (at least approximately),
and the corresponding effective level spacing is
$D\equiv-\del\log\Hml_\beta/\del\beta=\sigma_E^2/\Hml$.
It is then straightforward to write down the improved expression
for the statistical weight.
Moreover, the associated thermal distortion of the internal structure
of the wave packet was determined.

Since our initial suggestion \cite{Thermal} that this treatment might be useful
has led to some debate \cite{SF,OH},
we discussed and illustrated the various possible approaches to
determining the statistical behavior of one-body observables,
such as the occupation number.
In this manner, it was brought out that although
it is possible to recover the quantal appearance of single-particle observables
by suitable redefinition of the temperature parameter,
such attempts do not yield the proper many-body properties,
as governed by the behavior of the partition function.
The key to resolving the issue
lies in the inevitable distortion of the wave packet caused by
the canonical operator $\exp(-\beta\hat H)$.

To illustrate the practical utility of the treatment,
we considered the statistical properties of finite nuclei,
as exemplified by $^{12}$C and $^{40}$Ca.
The low-temperature behavior matches well with the 
finite-temperature liquid-drop model.
Moreover,
as the temperature is raised,
that the nuclear liquid drop evolves into a fragment gas,
as is expected for actual nuclar systems
and in good agreement with the standard statistical multifragmentation model.

We then turned to the important issue of how to incorporate
the quantum fluctuations into the dynamics.
Here the key suggestion is to allow the system to explore its
spectral distribution by introducing suitable stochastic transitions
between the wave packets \cite{PRL}.
This can conveniently be done by means of a Langevin term
in the equations of motion for the wave packet parameters.
We derived the general form of the associated transport coefficients
and verified that the proper microcanonical equilibrium distribution
is indeed achieved.
Simple approximate expressions
for the specific values of the transport coefficients
were then obtained by means of Fermi's golden rule,
leading to a practically useful treatment.
It is important to recognize that although the Langevin term
may resemble the effect of the standard collision term,
its origin is different:
While the former results from the residual interaction,
the latter arises from the quantum fluctuations 
that are inherent in the wave packets employed in the description.

The proposed extension of the standard treatment
represents a formally well based approach towards incorporating
the effect of quantum fluctuations into the wave packet dynamics.
Moreover,
it leads to the desired statistical properties in static scenarios
and can be included in the dynamics in a conceptually simple
and tractable manner.
The method may therefore find useful application
in the context of microscopic simulations of actual many-body processes,
such as fragment production in heavy-ion collisions.

This work was supported in part by
the Grant-in-Aid for Scientific Research (No.\ 06740193)
from the Ministry of Education, Science and Culture, Japan,
and by the Director, Office of Energy Research,
Office of High Energy and Nuclear Physics,
Nuclear Physics Division of the U.S. Department of Energy
under Contract No.\ DE-AC03-76SF00098.
%
%
One of the authors (A.O.) also thanks the Ministry of Education,
Science and Culture, Japan, for the Overseas Research Fellowship to him.
The calculations in this work were supported
by Research Center for Nuclear Physics (RCNP), Osaka University,
as RCNP Computational Nuclear Physics Project No.\ 93-B-03.

\appendix
\section{Moments of the transition rate}
\label{App}

In this appendix, we show some technical details regarding the calculation of
the transition rate and its first and second moments
which  are required in sect.\ \ref{VD}.

We start by rewriting the matrix element in the transition rate (\ref{rate}),
\beq
|\VEV{\Z'|\hat V|\Z}|^2
	=	{\cal V}(\ZC',\Z)\ {\cal V}(\ZC,\Z')\ 
		|\VEV{\Z'|\Z}|^2 \ ,
\eeq
where ${\cal V}(\ZC',\Z)\equiv(\Z'| \hat V |\Z)/(\Z'|\Z)$.
The overlap between two normalized states is approximately of Gaussian form,
\beq
\label{Overlap}
|\VEV{\Z'|\Z}|^2\ \approx\
		\exp\left(
			-\delta\ZC\cdot\C\cdot\delta\Z
		\right)
		\ ,
\eeq
as can be seen by expanding in $\delta\Z=\Z'-\Z$,
\beq
\log|\VEV{\Z'|\Z}|^2\ =\
	 -\delta\ZC\cdot\C\cdot\delta\Z\ +\ {\cal O}\left((\delta\z)^3\right)
		\ .
\eeq
The above expression (\ref{Overlap}) is consistent with the fact that
$d\ZC\cdot\C\cdot d\Z$ defines a infinitesimal squared distance
between two wave packets and,
accordingly, that $\det\C$ appears in the canonical measure.
Furthermore,
the relation is exact when $\C$ is unity,
{\em i.e.} when anti-symmetrization is ignored.

Using the expression (\ref{Overlap})
and assuming that $\Z'_\E$ is close to $\Z_\E$,
we obtain
\beqar
&&|\VEV{\Z'_\E| \hat V | \Z_\E}|^2\
\\
&\approx&\
	\left| {\cal V}(\ZC_\E,\Z_\E) \right|^2\ 
	\Exp{
		-\left(\ZC'_\E - \ZC_\E\right)
		\cdot\C\cdot
		\left(\Z'_\E - \Z_\E\right)
	}\ .
\eeqar
Finally,
when $\Z'_\E \approx \Z_\E$ then $\beta_{\Hml'}$ is small
and the distortion can be ignored, $\Z' \approx \Z'_\E$.
Thus we arrive at the total transition rate given in eq.\ (\ref{w0}).

We have used the resolution of unity,
\beqar
\nonumber
\int d\Gamma'\ \VEV{\Z|\Z'}\ \VEV{\Z'|\Z}\
	&=&\ 
\int d\Gamma'\ \Exp{-\delta\ZC\cdot\C\cdot\delta\Z}
\\
	&=& 1\ ,
\eeqar
and the first and second moment can be calculated by invoking the formulas
\beqar
&&
\int d\Gamma'\ \delta\z_n \exp\left(\delta\ZC\cdot\C\cdot\delta\Z\right) 
	= 0\ ,\\
&&
\int d\Gamma'\ \delta\z_n \delta\zc_{n'}
\exp\left(\delta\ZC\cdot\C\cdot\delta\Z\right) 
	= (C^{-1})_{nn'} \ .
\eeqar

In our previous work \cite{PRL},
we have subtracted the self-transition matrix element
and expanded ${\cal V}$ in $\delta\Z$.
While this subtraction reduces the total transition rate,
the relation between the drift and diffusion coefficients (\ref{Einstein})
remains valid.



\widetext

\begin{thebibliography}{99}
\bibitem{GSI}
        P. Kreutz et al., 
                \BIB{Nucl. Phys.}{A556}{1993}{672}.
\bibitem{Texas}
        K. Hagel et al.,
                \BIB{Phys. Rev. Lett.}{68}{1992}{2141}.
\bibitem{MSU}
        M.B. Tsang et al.,
                \BIB{Phys. Rev. Lett.}{71}{1993}{1502}.
\bibitem{BUU}
        G.F. Bertsch and S. Das Gupta,
                \BIB{Phys. Rep.}{160}{1988}{189};
        W. Cassing, V. Metag, U. Mosel, and K. Niita, 
                \BIB{Phys. Rep.}{188}{1990}{363}.
\bibitem{Aichelin}
        J. Aichelin and H. St\"ocker,
                \BIB{Phys. Lett.}{B176}{1986}{14};
        J. Aichelin,
                \BIB{Phys. Rep.}{202}{1991}{233};
        G. Peilert H. St\"ocker, W. Greiner,
                A. Rosenhauer, A. Bohnet, and J. Aichelin,
                \BIB{Phys. Rev.}{C39}{1989}{1402}.
\bibitem{Maruyama}
        Toshiki Maruyama, A. Ohnishi, and H. Horiuchi,
                \BIB{Phys. Rev.}{C42}{1990}{386};
                \BIB{Phys. Rev.}{C45}{1992}{2355};
        Toshiki Maruyama, A. Ono, A. Ohnishi, and H. Horiuchi, 
                \BIB{Prog. Theor. Phys.}{87}{1992}{1367}.
\bibitem{Boal}
        D.H. Boal and J.N. Glosli,
                \BIB{Phys. Rev.}{C38}{1988}{2621}.
\bibitem{AMD}
        A. Ono, H. Horiuchi, Toshiki Maruyama, and A. Ohnishi,
                \BIB{Phys. Rev. Lett.}{68}{1992}{2898}:
                \BIB{Prog. Theor. Phys.}{87}{1992}{1185};
                \BIB{Phys. Rev.}{C47}{1993}{2652}.
\bibitem{AMD2}
        A. Ono, H. Horiuchi, and Toshiki Maruyama,
                \BIB{Phys. Rev}{C48}{1993}{2946};
        A. Ono and H. Horiuchi,
                \BIB{Phys. Rev.}{C51}{1995}{299};
        E.I. Tanaka, A. Ono, H. Horiuchi, Tomoyuki Maruyama, and A. Engel,
                \BIB{Phys. Rev.}{C52}{1995}{316};
        A. Engel, E. I. Tanaka, Tomoyuki Maruyama, A. Ono, and H. Horiuchi,
                \BIB{Phys. Rev.}{C}{1995}{ to be published}.
\bibitem{FMD}
        H. Feldmeier,
                \BIB{Nucl. Phys.}{A515}{1990}{147};
        H. Feldmeier, K. Bieler, and J. Schnack,
                \BIB{Nucl. Phys.}{A586}{1995}{493};
        H. Feldmeier and J. Schnack,
                \BIB{Nucl. Phys.}{A583}{1995}{347}.
\bibitem{Thermal}
        A. Ohnishi and J. Randrup,
                \BIB{Nucl. Phys.}{A565}{1993}{474}.
\bibitem{Ono}
        A. Ono, private communication, 1992.
\bibitem{SF}
        J. Schnack and H. Feldmeier,
                preprint, GSI-95-34, 1995.
\bibitem{OH}
        A. Ono and H. Horiuchi,
                preprint, RIKEN-AF-NP-214, 1995.
\bibitem{PRL}
        A. Ohnishi and J. Randrup,
                \BIB{Phys. Rev. Lett.}{75}{1995}{596}.
\bibitem{Abe}
        Y. Abe, C. Gregoire, and H. Delagrange, 
                \BIB{J. Physique}{47}{1986}{C4-329};
        T. Wada, Y. Abe and N. Carjan,
                \BIB{Phys. Rev. Lett.}{70}{1993}{3538}.
\bibitem{Randrup}
        J. Randrup and B. Remaud,
		\BIB{Nucl. Phys.}{A514}{1990}{339};
        G.F. Burgio, Ph. Chomaz, and J. Randrup,
                \BIB{Nucl. Phys.}{A529}{1991}{157};
	F. Chapelle, G.F. Burgio, Ph. Chomaz, and J. Randrup,
		\BIB{Nucl. Phys.}{A540}{1992}{227}.
\bibitem{Ayik}
        S. Ayik and C. Gregoire,
                \BIB{Phys. Lett.}{B212}{1988}{269};
                \BIB{Nucl. Phys.}{A513}{1990}{187}.
\bibitem{Suraud}
        F.S. Zhang and E. Suraud, 
                \BIB{Phys. Lett.}{B319}{1993}{35}.
\bibitem{Volkov}
        A. Volkov,
                \BIB{Nucl. Phys.}{75}{1965}{33}.
\bibitem{Gross}
        D. H. E. Gross,
                \BIB{Rep. Prog. Phys.}{53}{1990}{605},
                \BIB{Nucl. Phys.}{A553}{1993}{175c};
        X.-Z. Zhang, et al.,
                \BIB{Nucl. Phys.}{A461}{1987}{641, 668}.
\bibitem{Fai}
        G. Fai and J. Randrup,
                \BIB{Nucl. Phys.}{A381}{1982}{557}.
\bibitem{Risken}
        Risken, {\em The Fokker-Planck Equation},
Springer (New York, 1989).
\end{thebibliography}
\end{document}